%% file: superstring-measure-genus-3.tex
\documentclass[11pt,a4paper]{article}
\pdfoutput=1 
\usepackage{jheppub}
\usepackage{tikz-cd}

\renewcommand\phi\varphi
\renewcommand\epsilon\varepsilon
\renewcommand\leq\leqslant
\renewcommand\geq\geqslant
\renewcommand\tilde{\widetilde}
\renewcommand\bar{\overline}

\title{RNS superstring measure for genus 3}
\author[a,d,e]{P. Dunin-Barkowski,}
\author[a,b,c,1]{I. Fedorov,\note{Corresponding author.}}
\author[e,f,g]{A. Sleptsov}

\affiliation[a]{Faculty of Mathematics, HSE University,\\
Usacheva 6, 119048 Moscow, Russia}

\affiliation[b]{International Laboratory of Cluster Geometry, Faculty of Mathematics, HSE University,\\
Usacheva 6, 119048 Moscow, Russia}

\affiliation[c]{Skolkovo Institute of Science and Technology (Skoltech),\\
Bolshoy Boulevard 30 bld. 1, 121205 Moscow, Russia}

\affiliation[d]{HSE--Skoltech International Laboratory of Representation Theory and Mathematical Physics, Skoltech,\\
Bolshoy Boulevard 30 bld. 1, 121205 Moscow, Russia}

\affiliation[e]{NRC ``Kurchatov Institute'',\\
123182 Moscow, Russia\footnote{Former Institute for Theoretical and Experimental Physics, 117218 Moscow, Russia.}}

\affiliation[f]{Institute for Information Transmission Problems,\\
127051 Moscow, Russia}

\affiliation[g]{Moscow Institute of Physics and Technology,\\
141700 Dolgoprudny, Russia}

\emailAdd{ptdunin@hse.ru}
\emailAdd{igoron-27@ya.ru}
\emailAdd{sleptsov@itep.ru}

\abstract{We propose a new formula for the RNS superstring measure for genus 3. Our derivation is based on invariant theory. We follow Witten's idea of using an algebraic parametrization of the moduli space (which he applied to re-derive D'Hoker and Phong's formula for the RNS superstring measure for genus 2); but the particular parametrization that we use has not been applied to superstring theory before. We prove that the superstring measure is a linear combination (with complex coefficients) of three known functions. Furthermore, we conjecture the values of the coefficients of this linear combination and provide evidence for this conjecture. Unlike the Ansatz of Cacciatori, Dalla Piazza and van Geemen from 2008, our formula has a polar singularity along the hyperelliptic locus; the existence of this singularity was established by Witten in 2015. Moreover, our formula is not an Ansatz but follows from first principles, except for the values of the three coefficients.}


\arxivnumber{2505.02950}

\begin{document}
\maketitle
\flushbottom

\section{Introduction}

We start with a historical introduction. The results of the present paper are summarized in section~\ref{sec:intro:new}.

\subsection{String amplitudes and the Mumford form}\label{sec:intro:mumford}

It is well known that path integrals of bosonic string theory can be reduced, via the Faddeev-Popov trick and a suitable regularization procedure, to integrals over finite-dimensional orbifolds, see e.g.\ \citep{D'HoPho88}. For the genus $g$ contribution to the vacuum amplitude ($g=2,3,...$), the domain of integration is the moduli space $\mathcal M_g$ of Riemann surfaces of genus $g$, and the integrand is known as the Polyakov measure $\Pi_g$. (In the case $g=1$ one chooses a marked point and works with $\mathcal M_{1,1}$ instead of $\mathcal M_1$, see \citep[section~2.1]{Witten13h}.\footnote{Witten does not make a distinction between $\mathcal M_1$ and $\mathcal M_{1,1}$, cf.\ \citep[appendix A]{Witten13h}. From our point of view, $\mathcal M_g$ for $g\geq2$ and $\mathcal M_{1,1}$ are orbifolds (i.e.\ Deligne-Mumford stacks), while $\mathcal M_1$ is something more complicated, because the automorphism group of any genus $1$ Riemann surface is infinite (translations of the torus). Fixing a point makes the automorphism group finite.} Throughout the present paper we only consider closed Riemann surfaces, and by default they are non-singular.)

Belavin and Knizhnik proved (\citep{BelKni86b, BelKni86a}, see also \citep{Voronov90} and \citep{BeiMan86}) that $\Pi_g$ is the ``modulus squared'' of a holomorphic quantity $\phi_g$, which is now known as the Mumford form. It had appeared in the mathematical literature almost a decade prior to that without any relation to string theory as a trivialization of a certain line bundle on $\mathcal M_g$ \citep[theorem 5.10]{Mumford77}. (To be more precise, Mumford proved that some trivialization exists; one can prove that if a trivialization exists, then it is unique up to a constant factor \citep[lemma 2.1]{HainReed02}.)

To compute string scattering amplitudes, it is desirable to have explicit formulas for $\Pi_g$. The most explicit formulas for $\phi_g$, and thence for $\Pi_g$, have been obtained when $g=1$ \citep{Shapiro72}, $2$ and $3$ \citep{BKMP86, BelKni86b, Morozov86} and a little less explicit formula (involving a residue) for $g=4$ \citep{BelKni86b, Morozov86, GuiMun95}. There are some formulas for $\phi_g$ valid for any $g$, e.g.\ 
\citep{BeiMan86, VerVer87, Ichikawa18}, but they are considerably less explicit.

In this paper we do not specify the normalizations, so we consider $\phi_g$ (and, consequently, $\Pi_g$) as defined up to a constant factor, as it is done in most papers we have referred to in this subsection.

\subsection{Superstring amplitudes and the super Mumford form}\label{sec:intro:super-mumford}

There is an analogous picture in type II RNS superstring theory: the path integral for the vacuum amplitude leads to the ``modulus squared'' of a holomorphic quantity $\psi_g$, called the super Mumford form \citep{RSchV89}. There is an alternative algebro-geometric definition of $\psi_g$ as a trivialization of a certain line bundle on the moduli space of super Riemann surfaces  \citep{Voronov88, RSchV89}. For computation of superstring scattering amplitudes, it is desirable to have explicit formulas for $\psi_g$.

The moduli space of super Riemann surfaces of given genus $g\geq 1$ has 2 connected components: $\mathcal S_g^-$ that corresponds to odd spin structures and $\mathcal S_g$ that corresponds to even ones. From now on, we shall focus on even spin structures, that is, on the component $\mathcal S_g$. The part of $\psi_g$ that lives over $\mathcal S_g^-$ is also important, but not that much; for example, it does not contribute to the vacuum amplitude (although it does contribute to \emph{some} amplitudes), cf.\ \citep{Witten13h}. So, from now on, we shall forget about $\mathcal S_g^-$ and assume that $\psi_g$ is defined on $\mathcal S_g$ when $g\geq 2$.

The supermoduli spaces $\mathcal S_g$ ($g\geq2$) are superorbifolds of dimension $3g-3|2g-2$; the bosonic truncation of $\mathcal S_g$ (obtained by setting all odd coordinates to zero) is the $(3g-3)$-dimensional moduli space $\mathcal M_g^+$ of Riemann surfaces with an even spin structure. Forgetting the spin structure corresponds to a covering map $c:\mathcal M_g^+\to\mathcal M_g$ of degree $2^{g-1}(2^g+1)$, which is the number of even spin structures on any genus $g\geq 1$ Riemann surface. See \citep{CodViv17}.

When $g=1$, one still needs a marked point, and actually in this case there are no odd moduli when only even spin structures are considered (this is explained e.g.\ in \citep[section~3]{Witten13h}). We have decided to forget about odd spin structures, so for us $\psi_1$ is defined on $\mathcal M_{1,1}^+$, a $3$-sheeted covering of $\mathcal M_{1,1}$.

$\psi_g$ is canonically normalized, but we do not consider normalizations in this paper, so for us $\psi_g$ is defined up to a constant factor.

\subsection{An explicit formula for the super Mumford form for genus 1}
In the following we consider the Mumford forms $\phi_g$ for $g=1,2,3$ as known quantities, cf.\ section~\ref{sec:intro:mumford}.

An explicit formula for $\psi_1$ has been known from the start, cf.\ e.g.\ \citep[eq. (3.259a)]{D'HoPho88}:
up to a constant factor $\psi_1/\phi_1$ corresponds to the modular form
\begin{equation}\label{eq:Xi1}
\Xi^{(1)} = \theta^8\!\!\begin{bmatrix}0\\0\end{bmatrix}\theta^4\!\!\begin{bmatrix}1\\0\end{bmatrix}\theta^4\!\!\begin{bmatrix}0\\1\end{bmatrix}
\end{equation}
(of genus 1, weight 8 and level $\Gamma_1(1,2)$); 
the notation for theta functions is recalled in appendix~\ref{app:theta} and the precise meaning of ``corresponds to'' is explained in section~\ref{sec:orbifolds}.

Here we abuse the notation slightly: the Mumford form $\phi_1$ is a form on $\mathcal M_{1,1}$, but we use the same symbol $\phi_1$ to denote the form on $\mathcal M_{1,1}^+$ obtained as the pullback of the Mumford form along the covering map $\mathcal M_{1,1}^+\to \mathcal M_{1,1}$. Thus $\psi_1/\phi_1$ is defined on $\mathcal M_{1,1}^+$. In the following we shall use the symbol $\phi_g$ ($g\geq2$) in the analogous manner, for both the Mumford form on $\mathcal M_g$ and its pullback to $\mathcal M_g^+$.

\subsection{D'Hoker and Phong's formulas for the super Mumford form for genus 2}

Explicit formulas for $\psi_2$ were only obtained in the beginning of the 2000's by D'Hoker and Phong in a breakthrough series of papers, see their survey \citep{D'HoPho02} and specifically \citep{D'HoPho01-4}.

To derive the formulas, D'Hoker and Phong introduced a procedure $\pi_*$ of integrating out odd coordinates. This allowed them to split $\psi_2$ into 2 components: $\psi_2\Big|_{\mathcal M_2^+}$ (coming from terms in $\psi_2$ of degree $0$ in odd coordinates) and $\pi_*\psi_2$ (coming from degree 2 terms), both well defined globally on $\mathcal M_2^+$. Then D'Hoker and Phong derived explicit formulas for $\psi_2\Big|_{\mathcal M_2^+}$ and $\pi_*\psi_2$.

For genus $2$ the superperiod map defines a holomorphic projection $\pi:\mathcal S_2\to \mathcal M_2^+$ from $\mathcal S_2$ to its bosonic truncation. In mathematical terms, $\pi_*$ is the integration along the fibres of $\pi$. 

Explicitly, D'Hoker and Phong's formula for $\pi_*\psi_2$ is as follows \citep[section~8]{D'HoPho02}: the form $\dfrac{\pi_*\psi_2}{\phi_2}$ extends holomorphically to the whole Siegel upper half-space $H_2$ as a genus $2$ Siegel modular form of the appropriate level and weight (it is clear a priori that the level should be $\Gamma_2(1,2)$ and the weight should be equal to $8$, see section~\ref{sec:orbifolds}), and this modular form, up to a constant factor, is
\begin{equation}\label{eq:Xi2}
\Xi^{(2)} = \theta^4\!\!\begin{bmatrix}00\\00\end{bmatrix}
\left(\!
\theta^4\!\!\begin{bmatrix}00\\11\end{bmatrix}
\theta^4\!\!\begin{bmatrix}01\\00\end{bmatrix}
\theta^4\!\!\begin{bmatrix}10\\01\end{bmatrix}
 +
\theta^4\!\!\begin{bmatrix}00\\01\end{bmatrix}
\theta^4\!\!\begin{bmatrix}01\\10\end{bmatrix}
\theta^4\!\!\begin{bmatrix}11\\00\end{bmatrix}
 +
\theta^4\!\!\begin{bmatrix}00\\10\end{bmatrix}
\theta^4\!\!\begin{bmatrix}10\\00\end{bmatrix}
\theta^4\!\!\begin{bmatrix}11\\11\end{bmatrix}
\right);
\end{equation}
the meaning of ``extends to'' is explained in section~\ref{sec:orbifolds}. This is the formula \citep[eq. (1.3)]{D'HoPho01-4} divided by \citep[eq. (7.14)]{D'HoPho01-4}; we have substituted $\delta=\begin{bmatrix}00\\00\end{bmatrix}$ (the meaning of this substitution is explained at the end of section~\ref{sec:orbifolds:hodge}) and an explicit expansion of \citep[eq. (1.5)]{D'HoPho01-4}. We also divided by $-\dfrac{\pi^6}{16}$: we choose the normalization in such a way that eq.~\eqref{eq:factorization-genus-2} below holds with $\Xi^{(1)}$ given by eq.~\eqref{eq:Xi1}.\footnote{What we denote $\Xi^{(2)}$ is denoted $\Xi_8[{}^{00}_{00}]$ in \citep{CDvG08}; they give 2 expressions for this modular form at the end of section~3, which actually differ by a sign. Our formula coincides with their 1st variant and with the negative of their 2nd variant.}

$\psi_1$, $\pi_*\psi_2$ and their higher-genus analogues are called ``(chiral) superstring measures'' in the literature, e.g.\ in \citep{D'HoPho02} and \citep{CDvG08}. In this paper we use the term ``superstring measure'' to refer to $\psi_1$, $\pi_*\psi_2$ or $\pi_*\psi_3$, where $\pi_*$ is the integration along the fibres of the superperiod map. (For genus $1$ there are no odd coordinates, so $\pi_*$ would not change anything for genus 1.)

\subsection{The Ansatz of Cacciatori, Dalla Piazza and van Geemen for genus 3}\label{sec:intro:ansatz}

In \citep{CDvG08, DvG08} the authors observed that $\Xi^{(2)}$ is the unique holomorphic Siegel modular form (of genus $2$, level $\Gamma_2(1,2)$ and weight 8) satisfying the following \emph{factorization condition:} 
\begin{equation}\label{eq:factorization-genus-2}
\Xi^{(2)}\begin{pmatrix}\tau'&0\\0&\tau''\end{pmatrix}=\Xi^{(1)}(\tau')\Xi^{(1)}(\tau'')
\end{equation}
with $\Xi^{(1)}$ given by~\eqref{eq:Xi1}.

They then tried to find a holomorphic Siegel modular form (of level $\Gamma_3(1,2)$ and weight $8$) satisfying the analogous factorization condition for genus $3$, i.e.\ coinciding with $\Xi^{(1)}(\tau_{11})\Xi^{(2)}\begin{pmatrix}\tau_{22}&\tau_{23}\\\tau_{23}&\tau_{33}\end{pmatrix}$ at block-diagonal matrices $\tau=\begin{pmatrix}\tau_{11}&0&0\\0&\tau_{22}&\tau_{23}\\0&\tau_{23}&\tau_{33}\end{pmatrix}$. 
And indeed they could prove that such a modular form exists, is unique and enjoys some other properties that one would expect from a genus 3 analogue of $\psi_1/\phi_1$ and $\pi_*\psi_2/\phi_2$ on physical grounds \citep{CDvG08, DvG08}.

It was also proved that holomorphic modular forms satisfying the genus $g$ factorization condition exist when $g=4$ and $5$. Some of these modular forms were observed to meet other expectations coming from superstring theory, and these forms were suggested as Ans\"atze (i.e.\ conjectural formulas) for the superstring measure. A review of this research direction can be found in \citep{DSS12}; here we only note that for genus $4$ all proposed Ans\"atze coincide with the one proposed by Grushevsky in \citep{Grushevsky08}.

The survey \citep{Morozov08} revisits what was known about string and superstring measures in 2008.

\subsection{Criticisms of the Ansatz}

Some problems with all these Ans\"atze for genera $g\geqslant3$ were indicated later in the literature.

In \cite{DMS09} it was noticed that none of the proposed Ans\"atze could work for genus $6$, and certain problems with the Ans\"atze for genus $5$ were also indicated. Some of the problems were then fixed in \cite{DSS12} for genus 5, but not for genus 6.

Later on it was indicated that the very interpretation of the existent Ans\"atze for $g\geq3$ was problematic. It is not clear in what exact way the Ans\"atze (modular forms of genera $g=3,4,5$) should possibly be related to $\psi_g$ and thus to superstring theory. The Ans\"atze were being derived essentially by axiomatizing some properties of D'Hoker and Phong's
modular form $\Xi^{(2)}$ describing $\pi_*\psi_2/\phi_2$, where $\pi_*$ is the integration along the fibres of the superperiod map. But is there a natural analogue of $\pi_*\psi_2/\phi_2$ for higher genera?

First of all, the superperiod map does not define a projection from $\mathcal S_g$ to its bosonic truncation $\mathcal M_g^+$ for any $g\geq 4$ (not even a meromorphic one): the image of the superperiod map is non-reduced when $g\geq 4$ \citep[remark 6.8]{CodViv17}, so it cannot be a piece of $\mathcal M_g^+$; rather, it is an ``infinitesimal thickening'' of an open and dense piece of $\mathcal M_g^+$. The description of this infinitesimal thickening is the superversion of the Riemann-Schottky problem \citep{CodViv17, FKP19}. So $\pi_*\psi_g$ is undefined for $g\geq 4$, at least if $\pi_*$ should stand for the integration along the fibres of the superperiod map. This problem is discussed briefly in \citep[the end of section 3]{Witten13h}. It is not clear how to define fibrewise integration in such a context, when $\pi$ is not a projection but something more complicated.

Another objection refers to genus $3$. The superperiod map does define a meromorphic projection $\pi:\mathcal S_3\to \mathcal M_3^+$ in this case, although this projection is not everywhere holomorphic but has poles over the hyperelliptic divisor $\mathcal H_3\subset \mathcal M_3$ \citep[appendix C.3]{Witten15} (cf.\ \citep[theorem 6.3]{BHP20}). So for genus $3$ one may still consider the form $\pi_*\psi_3$, but a priori it is only well defined outside of the locus of hyperelliptic curves, while it may in principle have poles over $\mathcal H_3$.

In \citep[appendix C.4]{Witten15} Witten showed that $\pi_*\psi_3$ does indeed have a pole and computed the order of the pole.\footnote{In an earlier preprint Witten stated without proof that $\pi_*\psi_3$ should be holomorphic on $\mathcal M_3^+$ \citep[the end of section 3]{Witten13h}. The results of \citep[appendix C.4]{Witten15} refute that earlier statement.} $\phi_3$ has no poles and no zeros, so $\dfrac{\pi_*\psi_3}{\phi_3}$ has a pole too (of the same order as $\pi_*\psi_3$). On the other hand, the Ansatz of \citep{CDvG08} is holomorphic everywhere on the Siegel upper half-space $H_3$, so this Ansatz cannot be a formula for $\dfrac{\pi_*\psi_3}{\phi_3}$. {Indeed, a holomorphic modular form on $H_3$ of level $\Gamma_3(1,2)$ describes a holomorphic section of a line bundle on $\mathcal A_3^+ = H_3/\Gamma_3(1,2)$, see section~\ref{sec:orbifolds}; the period map $\mathcal M_3^+\to \mathcal A_3^+$ is holomorphic, so it pulls back holomorphic sections of line bundles on $\mathcal A_3^+$ to holomorphic sections of line bundles on $\mathcal M_3^+$.\footnote{Note that this argument does not work in the inverse direction: a holomorphic section of a line bundle on $\mathcal M_3^+$ or $\mathcal M_3$ need not extend to a holomorphic modular form. For example, the Mumford form $\phi_3$ is holomorphic on $\mathcal M_3$, and so the Polyakov measure $\Pi_3 = |\phi_3|^2$ is non-singular everywhere on $\mathcal M_3$, notwithstanding that $\Pi_3$ is described by a function on $H_3$ that has a polar singularity along the hyperelliptic locus \citep{BKMP86}.}
}

In principle, there remains a possibility that the Ansatz of \citep{CDvG08} describes $\dfrac{\tilde\pi_*\psi_3}{\phi_3}$ for some other projection $\tilde\pi:\mathcal S_3\to \mathcal M_3^+$. As of now, no one has constructed such a $\tilde\pi$. It is not known whether a holomorphic projection $\mathcal S_g\to \mathcal M_g^+$ exists at all for $g=3$ or $4$, while it is known that such a projection does not exist for any $g\geq 5$ \citep{DonWit13}. (One may think that sending each super Riemann surface to its underlying Riemann surface with a spin structure is a holomorphic projection $\mathcal S_g\to \mathcal M_g^+$ for any $g$. But in fact this does not define a map $\mathcal S_g\to \mathcal M_g^+$: it is not enough to specify what the map does at the level of points to define a map of supermanifolds.)

Note that $\psi_g$ is not just a Berezinian volume form but a Berezinian volume form valued in a line bundle, namely, in the bundle $b^{-5}$ (see section~\ref{sec:derivation:bundles}). Therefore in order to integrate $\psi_g$ along the fibres of a projection $\tilde\pi$ one would need not just $\tilde\pi$ itself but also some additional structure. An isomorphism of vector bundles $b^{-5}\to\tilde\pi^*\lambda^{-5}$ on $\mathcal S_g$ would certainly suffice, but such an isomorphism may in principle fail to exist even if $\tilde\pi$ exists. See section~\ref{sec:derivation:bundles} for some more details and references.

We note that, despite the problems that we have just indicated, the genus 3 Ansatz of \citep{CDvG08} formed the basis of a proposal for the complete genus 3 4-point amplitude in a recent work \citep{GMS21}, and their proposal matches some partial results obtained in \citep{GomMar13} via the pure-spinor formalism; see also point \ref{sec:conclusion:ansatz} in section~\ref{sec:conclusion} below.

\subsection{A new formula for genus 3}\label{sec:intro:new}

In the present paper we propose a new formula for $\pi_*\psi_3$ (\emph{the genus 3 superstring measure}, see remark~\ref{rem:SSMeasureName} at the end of the present section). We recall that $\psi_3$ is the super Mumford form restricted to the even component of the moduli space of super Riemann surfaces of genus 3, and $\pi_*$ is the integration along the fibres of the meromorphic projection $\pi$ (induced by the superperiod map) from the moduli space of super Riemann surfaces onto its bosonic truncation, the moduli space of Riemann surfaces with an even spin structure.

The formula for the Mumford form $\phi_3$ is known (see section~\ref{sec:intro:mumford}), so it is enough to derive a formula for the ratio $\dfrac{\pi_*\psi_3}{\phi_3}$. We write our formula for $\dfrac{\pi_*\psi_3}{\phi_3}$ in two ways.

First we prove that $\pi_*\psi_3/\phi_3$ is a linear combination of three explicitly known functions given in terms of invariant theory of nets of quaternary quadrics; the proof occupies sections~\ref{sec:derivation:bundles}--\ref{sec:derivation:18} (see points~\ref{sec:plan:bundles}--\ref{sec:plan:18} of our plan in section~\ref{sec:plan}):

\begin{equation}\label{eq:Xi3-invariants:intro}
\dfrac{\pi_*\psi_3}{\phi_3}=(k_1\Lambda^3+k_2I_3\Lambda+k_3Q')IJ\eta^8.
\end{equation}
Here $I,J,\Lambda, I_3$ and $Q'$ are particular invariants of nets (explicit formulas for these invariants are given in appendix~\ref{app:invariants}), $\eta$ is a certain standard trivialization of the Hodge bundle on the space of parameters and $k_1,k_2,k_3$ are three complex numbers that remain unknown at this step.

Then we rewrite our formula in terms of Siegel modular forms. This reformulation is partly conjectural, because at some point it relies on computer calculations which are convincing but not sufficient as a proof; our argument is given in section~\ref{sec:derivation:modular-forms} (see point~\ref{sec:plan:modular-forms} of our plan in section~\ref{sec:plan} for the explanation of the notation):

\begin{subequations}\label{eq:Xi3-modular:intro}
\begin{equation}\label{eq:Xi3-modular-1:intro}
\dfrac{\pi_*\psi_3}{\phi_3}=\Xi^{(3)}dz^{8},
\end{equation}
\begin{equation}\label{eq:Xi3-modular-2:intro}
\Xi^{(3)}(\tau)=\dfrac{\left(k_1\Lambda^3 + k_2I_3\Lambda + k_3Q'\right)\!I}{J}\Big(A(\tau)\Big)\;{\theta_{00}^{16}(\tau)}.
\end{equation}
\end{subequations}

$\Xi^{(3)}$ is a meromorphic Siegel modular form of genus 3, level $\Gamma_3(1,2)$ and weight $8$.

This reformulation via Siegel modular forms also allows us to conjecture the values of the three unknown parameters appearing in \eqref{eq:Xi3-invariants:intro} and~\eqref{eq:Xi3-modular-2:intro}; the evidence for this conjecture is given in section~\ref{sec:derivation:factorization} (see point~\ref{sec:plan:factorization} of our plan in section~\ref{sec:plan}):

\begin{equation}\label{eq:3-parameters}
\begin{aligned}
k_1&= 2^8 \cdot 3^7 \cdot 5^2 \cdot 7^2 \cdot 11^2 \cdot 13^2 \cdot 17 \cdot 19 \cdot 23,\\
k_2&=0,\\
k_3&=-2^2 \cdot 3^3 \cdot 5 \cdot 7^2 \cdot 11 \cdot 13 \cdot 23.
\end{aligned}
\end{equation}

Why these numbers end up being integers, and quite special ones at that, is not clear at the moment, see point~\ref{sec:conclusion:coefficients} in section~\ref{sec:conclusion} for a discussion.

\vspace{0.7em}
\textit{Remarks:}
\vspace{-0.7em}
\begin{enumerate}
\item We note that our formula~\eqref{eq:Xi3-invariants:intro} for $\pi_*\psi_3$ with three unknown parameters is derived from the first principles of the theory: we did not rely on any unproved properties or constraints to get this formula for $\pi_*\psi_3$. This distinguishes our approach from previous approaches to finding formulas for superstring measures for genus 3 \citep{D'HoPho04a, D'HoPho04b}, \citep{CDvG08} and higher. Our derivation of eq.~\eqref{eq:Xi3-modular-2:intro} is rigorous too (assuming that the relation~\eqref{eq:J5-over-I8} between modular forms holds). In contrast, the conjectured values of the three parameters~\eqref{eq:3-parameters} are derived not from the principles of superstring theory but from a version of the factorization condition, as in \citep{CDvG08}; this factorization condition is only a conjecture.

\item \label{rem:WittenPole} Our derivation does \emph{not} use Witten's result on the pole of $\pi_*\psi_3$ \citep[appendix C.4]{Witten15}. On the contrary, we can re-derive Witten's result as a corollary of our formula. With our technique we can also compute the order of the pole, and it coincides with the order computed by Witten. This is going to be treated in a future publication.

\item \label{rem:SSMeasureName} We call $\pi_*\psi_3$ the genus 3 superstring measure. The term \emph{superstring measure} has come to have different meanings in the literature. Sometimes it refers to the super Mumford form or its ``modulus squared'', defined on supermoduli spaces.
We, on the contrary, use the term \emph{superstring measure} in the same manner as the authors of \citep{CDvG08}: not for the whole super Mumford form but for some kind of projection thereof, defined on the (non-super) moduli space $\mathcal M_3^+$. The difference is that the authors of \citep{CDvG08} do not specify exactly what projection they use, while we make it certain that we use the projection $\pi_*$ induced by the superperiod map.

\item Note that both formulas~\eqref{eq:Xi3-invariants:intro} and~\eqref{eq:Xi3-modular-2:intro}, with numerical coefficients~\eqref{eq:3-parameters} substituted, are completely explicit once one substitutes all of their ingredients as we indicate in the present paper. For instance, after the substitution of all ingredients, formula~\eqref{eq:Xi3-modular-2:intro} becomes the quotient of two completely explicit, if bulky, polynomials (with rational coefficients) in theta constants $\theta\!\begin{bmatrix}a\\b\end{bmatrix}\!(0,\tau)$ and theta gradients $\dfrac{\partial}{\partial z_i}\Bigg|_{z=0}\theta\!\begin{bmatrix}a\\b\end{bmatrix}\!(z,\tau)$ (notation for theta functions is recalled in appendix \ref{app:theta}). This explicit expression is too bulky to be given directly in the paper; one can hope to make it much shorter by using various relations between theta constants and theta gradients, but this remains to be seen. We have made the expression available on the Internet via \url{https://github.com/igorf-27/superstring-measure-genus-3}.

\end{enumerate}

\subsection{Two main ideas}

Let us outline the two main ideas our derivation is based upon:

\begin{enumerate}
\item Algebraic parametrizations (following Witten). The entries $\tau_{ij}$ ($i\leq j$) of period matrices can be used as coordinates on $\mathcal M_3^+$. However, this parametrization is only analytic, not algebraic: the Siegel upper half-space is not an algebraic variety. On the other hand, everything else in sight is algebraic, in particular, $\pi_*\psi_3/\phi_3$ is, because $\phi_g$, $\psi_g$ and the superperiod map can be defined purely in terms of algebraic geometry without resorting to analytic techniques \citep{Mumford77, FKP20, BHP20}.

The idea is to use some algebraic parametrization of moduli spaces instead of the parametrization with the $\tau_{ij}$'s: then we'll have to search for rational functions (quotients of 2 polynomials), not for general meromorphic ones, and this will make the quest easier.

Witten used this idea in \citep{Witten13h} to perform an alternative derivation of D'Hoker and Phong's formula for $\pi_*\psi_2$. D'Hoker and Phong's derivation is complicated and relies on path integral techniques, while Witten's derivation is simpler, because he uses an algebraic parametrization.

\item The second main idea is the particular choice of parametrization. To our knowledge, this parametrization has not been applied to string theory before. To parametrize the moduli space of even spin genus 3 curves, we use the following theorem from classical algebraic geometry \citep[proposition 4.2]{Beauville99}. Let $f$ be a homogeneous polynomial of degree 4 in 3 variables $x_0,x_1,x_2$. Suppose that the curve $C$ in $\mathbb P^2$ defined by the equation $f=0$ is smooth. Then there is a natural bijection between even spin structures on $C$ and representations $f=\det A$, where $A$ is a symmetric $4\times4$ matrix with linear functions $A_{kl}=x_0A_{0kl}+x_1A_{1kl}+x_2A_{2kl}$ as entries ($A_{ikl}\in \mathbb C$), up to a natural action of the group $GL_3\times GL_4$ on the space of such matrices; more details below, in section~\ref{sec:derivation:determinantal}. We shall use $A_{ikl}$ as parameters. 
\end{enumerate}

Note that our work on the 3-loop superstring measure did not simply amount to choosing this parametrization and just following what Witten did in the genus 2 case. The parametrization in terms of $A_{ikl}$ is considerably more complicated than the hyperelliptic parametrization that Witten used for genus 2. We could not use the hyperelliptic parametrization for the genus 3 case, because hyperelliptic Riemann surfaces of genus 3 only form a subspace of codimension one in the moduli space of all Riemann surfaces of genus 3 (while any genus 2 Riemann surface is hyperelliptic). Moreover, the hyperelliptic locus in $\mathcal M_3^+$ has two irreducible components, and, curiously, the superstring measure is identically zero on one of the components and develops a pole along the other one \citep[Appendix C.4]{Witten15}. All in all, the case of genus 3 is more complicated and not really analogous to that of genus 2; there were quite a few significant new problems which we had to overcome.

\subsection{Testing certain implications of the formula}

Let us briefly discuss what implications of our formula for the genus 3 superstring measure we have been able to test so far.
\begin{enumerate}
	\item It is expected that the genus $g$ contribution to the \emph{0-point function} of type II superstring theory on flat 10-dimensional space vanishes for all $g$ \citep{Martinec86}. For the case $g=2$ this was confirmed by D'Hoker and Phong: their formula for the superstring measure $\pi_*\psi_2$ implies that the sum of $\pi_*\psi_2$ over all even spin structures (theta characteristics) vanishes at every point of the moduli space \citep[section 6.3]{D'HoPho01-4}, and this fact, combined with some general considerations as to the infrared behaviour of $\psi_2$ \citep[section 3.2]{Witten13p}, \citep[section 19]{D'Hoker14}, implies that the genus 2 0-point function vanishes.

In section~\ref{sec:vacuum:interior} we deduce from our formula \eqref{eq:Xi3-invariants:intro} that the sum of $\pi_*\psi_3$ over even spin structures vanishes too. In section~\ref{sec:vacuum:boundary} we discuss the relation of this fact to the vanishing of the genus 3 0-point function.	
	\item The conjectural factorization condition discussed in section~\ref{sec:derivation:factorization} does not only determine the values \eqref{eq:3-parameters} of the coefficients $k_1,\, k_2,\, k_3$ uniquely (section~\ref{sec:derivation:factorization:verification}). Let $H_g\subset\mathbb C^{g\times g}$ be the Siegel upper half space. The factorization condition also produces from our genus 3 formula \eqref{eq:Xi3-modular:intro} a function on $H_1$ and a function on $H_2$ (both defined up to a constant factor) that coincide, as numerical tests suggest, with the genus 1 and the genus 2 superstring measure. The latter was first computed by D'Hoker and Phong, but here we get it independently of their results. See section~\ref{sec:derivation:factorization:recovering}.  We believe this to be an important check of the validity of our formula.
	
	\item As mentioned in remark~\ref{rem:WittenPole} at the end of section~\ref{sec:intro:new}, by using our formula we can re-derive Witten's result on the order of the pole of $\pi_*\psi_3$ (which Witten presented in \citep[appendix C.4]{Witten15}). We have carried out the computations indicating this, but a careful discussion of this lies beyond the scope of the present paper and is relegated to a subsequent publication.
\end{enumerate}

\subsection{The structure of the paper}

Section~\ref{sec:plan} contains the detailed plan of the derivation of our new formula for $\pi_*\psi_3$.

In section~\ref{sec:orbifolds} we review some preliminary material about orbifolds. In particular, we explain the connection between the abstract definition of $\pi_*\psi_3$ as a section of a line bundle and Siegel modular forms.

In section~\ref{sec:derivation} we implement the points of the plan of section~\ref{sec:plan}.

In section~\ref{sec:vacuum} we deduce from our formula \eqref{eq:Xi3-invariants:intro} that the sum of $\pi_*\psi_3$ over spin structures vanishes. We also explain the relation between this result and the 0-point function of type II superstring theory.

In section~\ref{sec:conclusion} we summarize our results and indicate some questions for further research.

\section{The plan of the derivation}\label{sec:plan}

Let us give the plan of our derivation of the new formula for $\pi_*\psi_3$. The points of the plan are implemented in the respective subsections of section~\ref{sec:derivation}, which we number in exactly the same way.

\begin{enumerate}
\item\label{sec:plan:bundles}
We start with an abstract description of $\pi_*\psi_3$ as a section of a line bundle on the moduli space:
\begin{equation}
\pi_*\psi_3\in H^0\left(\mathcal M_{3,nh}^+,\omega_{\mathcal M_{3,nh}^+}\otimes\lambda^{-5}\right).
\end{equation}

Here $\mathcal M_{3,nh}^+\subset \mathcal M_3^+$ is the moduli space of non-hyperelliptic genus 3 Riemann surfaces with an even spin structure, $\omega_{\mathcal M_{3,nh}^+}$ is the bundle of holomorphic volume forms (i.e.\ $3g-3=6$-forms) on $\mathcal M_{3,nh}^+$ and $\lambda$ the Hodge line bundle.

\item As an explicit formula for $\phi_3$ is known (see section~\ref{sec:intro:mumford}), we choose to focus on the ratio
\begin{equation}
\dfrac{\pi_*\psi_3}{\phi_3}\in H^0\left(\mathcal M_{3,nh}^+,\lambda^8\right).
\end{equation}

This step is not very important, it just makes some formulas shorter.

\item We describe the algebraic parametrization of $\mathcal M_{3,nh}^+$ in terms of the parameters $A_{ikl}\in\mathbb C$ ($0\leq i\leq 2$, $0\leq k\leq l\leq 3$).

\item We study how to describe sections of the Hodge line bundle on $\mathcal M_3^+$ and of its tensor powers in terms of the chosen parametrization. It turns out that meromorphic sections of $\lambda^k$ correspond bijectively to $(SL_3\times SL_4)$-invariant rational functions of degree $12k$ on the space of parameters.

\item From the fact that $\pi_*\psi_3$ is regular on $\mathcal M_3^+$ outside of the hyperelliptic locus (and $\phi_3$ is regular and non-zero everywhere on $\mathcal M_3^+$) we obtain the formula
\begin{equation}
\dfrac{\pi_*\psi_3}{\phi_3}=PI^aJ^b\eta^8.
\end{equation}

Here $\eta$ is a certain standard trivialization of the Hodge bundle on the space of parameters, $I$ and $J$ are certain known $(SL_3\times SL_4)$-invariant polynomial functions of $A_{ikl}$ (explicit formulas are in appendix~\ref{app:invariants}), $a$ and $b$ are unknown integers and $P$ is an unknown invariant polynomial function of $A_{ikl}$.

\item By analyzing the behaviour of $\pi_*\psi_3/\phi_3$ at infinity, we determine $a=b=1$. It follows that the degree of $P$ is $18$.

\item\label{sec:plan:18}
We determine that the vector space of polynomial invariants of degree $18$ is 3-dimensional via a computer-assisted proof. In the literature we have found three linearly independent degree 18 invariants 
\begin{equation}
P_1=\Lambda^3,\quad
P_2=I_3\Lambda
\quad\text{and}\quad
P_3=Q',
\end{equation}
see \citep[section 5]{Gizatullin07} and appendix~\ref{app:invariants} for more details. Our result on the dimension then implies that $P_1,P_2,P_3$ constitute a basis of the space of invariants of degree 18.

At this point we have obtained the formula
\begin{equation}\label{eq:Xi3-invariants:plan}
\dfrac{\pi_*\psi_3}{\phi_3}=(k_1P_1+k_2P_2+k_3P_3)IJ\eta^8,
\end{equation}
where everything is known apart from the three complex parameters $k_1,k_2,k_3$.
\end{enumerate}

The following two points of the plan are partly conjectural.

\begin{enumerate}
\setcounter{enumi}{7}
\item\label{sec:plan:modular-forms}
We translate the description of $\dfrac{\pi_*\psi_3}{\phi_3}$ from the language of invariant theory into the language of modular forms. This translation is partly conjectural: we need to know that a certain relation~\eqref{eq:J5-over-I8} between modular forms holds. We have checked this relation numerically at a number of values of $\tau$ with a computer and observed that it holds for these values, but we do not have a complete proof.

In this way we get our formula for $\pi_*\psi_3/\phi_3$ in terms of Siegel modular forms:

\begin{subequations}
\begin{equation}
\dfrac{\pi_*\psi_3}{\phi_3} = \Xi^{(3)}  dz^8,
\end{equation}
\begin{equation}\label{eq:Xi3-modular-2:plan}
\Xi^{(3)}(\tau) = \dfrac{(k_1P_1 + k_2P_2 + k_3P_3)I}{J}(A(\tau))\;\theta_{00}^{16}(\tau).
\end{equation}
\end{subequations}
where the three complex parameters $k_1,k_2,k_3$ are still unknown.

Here $dz$ is the standard trivialization of the Hodge bundle on the Siegel upper half-space (see section~\ref{sec:orbifolds}), $\theta_{00}(\tau)$ is the theta constant with characteristic $\begin{bmatrix}000\\000\end{bmatrix}$ (see appendix~\ref{app:theta}) and $A(\tau)$ a certain meromorphic function on the Siegel upper half-space $H_3$ valued in the space of nets (an explicit formula for $A(\tau)$ is provided in appendix~\ref{app:map}). The fraction in this formula is a rational invariant of nets, and this invariant is being evaluated at the net $A(\tau)$. So $\Xi^{(3)}$ is a meromorphic function on the Siegel upper half-space; it is actually a meromorphic Siegel modular form of level $\Gamma_3(1,2)$ and weight $8$.

\item \label{sec:plan:factorization}
Now we want to impose an analogue of the factorization condition of \citep{CDvG08}. We note that $\Xi^{(3)}$ is undefined (has no limit) when $\tau$ is block-diagonal: $\tau=\begin{pmatrix}\tau_{11}&0&0\\0&\tau_{22}&\tau_{23}\\0&\tau_{23}&\tau_{33}\end{pmatrix}$. So it is impossible to substitute such a $\tau$ directly. We suggest a certain regularized substitution procedure and observe, relying on computer calculations, that there exists one and only one triple $(k_1,k_2,k_3)$ of complex numbers such that the regularized substitution of the block diagonal $\tau$ as above into the right-hand side of the formula~\eqref{eq:Xi3-modular-2:plan} for $\Xi^{(3)}(\tau)$ gives $\Xi^{(1)}(\tau_{11})\Xi^{(2)}\begin{pmatrix}\tau_{22}&\tau_{23}\\\tau_{23}&\tau_{33}\end{pmatrix}\!;$ this is the triple~\eqref{eq:3-parameters}. Conjecturally, this triple gives the right formula for $\Xi^{(3)}$.
\end{enumerate}

\section{Line bundles on orbifolds}\label{sec:orbifolds}

The aim of this section is to explain the relation between sections of line bundles on moduli spaces (such as $\pi_*\psi_3$) and coordinate formulas describing these abstract sections in terms of some parametrization of the moduli space.

The main point is that one can use different parametrizations to describe the same section, and sometimes one parametrization is more convenient than another. For example, formulas for string and superstring measures have usually been written in coordinates given by period matrices. In contrast, in section~\ref{sec:derivation} we use a different parametrization of the spin moduli space in terms of determinantal representations of quartics. It is this parametrization that enables us to derive the formula for $\pi_*\psi_3$.

The parametrization via period matrices leads to formulas in terms of modular forms, whereas our parametrization leads to formulas in terms of invariant theory. The description via modular forms and the description via invariants can be converted one into another as long as they refer to the same section of a line bundle on the moduli space.

\subsection{Sections of a line bundle on an orbifold as functions on the space of parameters}\label{sec:orbifolds:bundles}

Let $U\subset\mathbb C^n$ be a domain with an action of a Lie group $G$ by biholomorphic automorphisms such that $M=U/G$ is an orbifold. 
A line bundle $L$ on $M$ is the same thing as a line bundle $\tilde L$ on $U$ with a fibrewise linear action of $G$ on the total space of $\tilde L$ that extends the action of $G$ on $U$. Let $t$ be a trivialization of $\tilde L$ (a globally defined holomorphic section with no zeros); then $gt$ is also a trivialization for any $g\in G$, so $gt=e_gt$ for a nowhere zero holomorphic function $e_g$ on $U$. The collection $e=\{e_g|g\in G\}$ is called the \emph{automorphy factor} or the \emph{multiplier system} corresponding to $t$. Now, if $s$ is a holomorphic (meromorphic) section of $L$, then $\theta:=s/t$ is a holomorphic (meromorphic) function on $U$ with the property $\theta(gx)=e_g(x)^{-1}\theta(x)$; conversely, any such function $\theta$ defines a section of $L$. In this situation we write
\begin{equation}
s=\theta t.
\end{equation}

What we call an ``explicit coordinate formula for $s$'' is an explicitly given holomorphic function $\theta$ on $U$ that describes $s$, for some given presentation $M=U/G$ and trivialization $t$.

\subsection{The Hodge line bundle and modular forms}\label{sec:orbifolds:hodge}

Here we recall some standard facts about moduli spaces; cf.\ e.g.\ \citep[section 2]{Farkas12} and \citep{CDvG08}.

For a fixed genus $g\geq2$ let us consider the following orbifolds: the moduli space $\mathcal M_g$ of Riemann surfaces; its covering $\mathcal M_g^+$, the moduli space of Riemann surfaces with an even spin structure; the moduli space $\mathcal A_g$ of principally polarized Abelian varieties; and its covering $\mathcal A_g^+$, the moduli space of principally polarized Abelian varieties with an even theta characteristic. For $g=1$ we don't consider $\mathcal M_1$ nor $\mathcal M_1^+$, as they are not orbifolds; instead we consider $\mathcal M_{1,1}=\mathcal A_1$ and $\mathcal M_{1,1}^+=\mathcal A_1^+$.

$\mathcal A_g=H_g/\Gamma_g$ and $\mathcal A_g^+=H_g/\Gamma_g(1,2)$, where $H_g\subset\mathbb C^{g\times g}$ is the Siegel upper half-space, $\Gamma_g=Sp_{2g}(\mathbb Z)$, $\Gamma_g(1,2)\subset\Gamma_g$ is the Igusa subgroup and $M=\begin{pmatrix}A&B\\C&D\end{pmatrix}\in\Gamma_g$ maps $\tau\in H_g$ to $(A\tau+B)(C\tau+D)^{-1}$.  The classical period map $\mathcal M_g\to \mathcal A_g$ lifts to the holomorphic map $\mathcal M_g^+\to \mathcal A_g^+$.

With the letter $\lambda$ we denote the Hodge line bundle on any of these moduli spaces. For $\mathcal M_g$ (or $\mathcal M_g^+$), the fibre of $\lambda$ at a Riemann surface $C\in \mathcal M_g$ is the 1-dimensional complex vector space $\lambda_C=\bigwedge^gH^0(C,\omega_C)$; here $\omega_C=\Omega^1_C$ denotes the line bundle of holomorphic 1-forms on $C$ and $H^0(C,\omega_C)$ its space of global sections. It is well known that the dimension of the vector space of global holomorphic 1-forms on a Riemann surface of genus $g$ is precisely $g$, so the $g$'th exterior power $\lambda_C$ is indeed 1-dimensional. Analogously, for $\mathcal A_g$ (or $\mathcal A_g^+$), the fibre of $\lambda$ at a complex torus $J\in \mathcal A_g$ is the 1-dimensional complex vector space $\lambda_J=\bigwedge^gH^0(J,\Omega^1_J)$. If $J$ is the Jacobian of $C$, then $\lambda_J$ and $\lambda_C$ are canonically isomorphic, so the pullback of the Hodge bundle from $\mathcal A_g$ (or $\mathcal A_g^+$) can be identified with the Hodge bundle on $\mathcal M_g$ (or $\mathcal M_g^+$), that is why we denote them with the same letter.

The complex torus over $\tau\in H_g$ is $\mathbb C/(\mathbb Z^g\oplus\tau\mathbb Z^g)$. The 1-form $dz_i$ on $\mathbb C^g$ is invariant under translations, so it descends to the torus, and we can choose $dz:=dz_1\wedge...\wedge dz_g$ as a trivialization of the Hodge bundle on $H_g$. The corresponding automorphy factor is $e_M(\tau)=\det(C\tau+D)^{-1}$ (see \citep[p. 141]{FalChai}); so holomorphic sections of $\lambda^d$ on $\mathcal A_g$ (resp. $\mathcal A_g^+$) correspond bijectively to holomorphic functions $f:H_g\to\mathbb C$ such that 
\begin{equation}\label{eq:modularity}
f((A\tau+B)(C\tau+D)^{-1})=\det(C\tau+D)^df(\tau)
\end{equation}
for any $M=\begin{pmatrix}A&B\\C&D\end{pmatrix}\in\Gamma_g$ (resp. $\Gamma_g(1,2)$). When $g\geq2$, such functions are called Siegel modular forms of genus $g$, weight $d$ and level $\Gamma_g$ (resp. $\Gamma_g(1,2)$). One can change ``holomorphic'' to ``meromorphic'' in this paragraph, then everything will remain true and one will get so called meromorphic Siegel modular forms.

When $g\geq2$, every meromorphic section of $\lambda^d$ is rational, i.e.\ comes from algebraic geometry. For $g=1$ this is not the case. Meromorphic modular forms of weight $d$ and level $\Gamma$ for $g=1$ are defined as those functions $f:H_1\to\mathbb C$ that correspond to rational sections of $\lambda^d$ (not just to meromorphic ones); this means that in addition to the transformation property~\eqref{eq:modularity} $f$ must be ``meromorphic at the cusps of $\Gamma$''.\footnote{For $\Gamma = \Gamma_1$ ``$f$ is meromorphic at the cusps'' means that $f$ has no essential singularity at $q=0$ as a function of $q:=\exp(\pi i\tau)$. For $\Gamma=\Gamma_1(1,2)$ this condition means that both $f$ and the function $g(\tau):=f\left(\dfrac{\tau}{\tau+1}\right)$ have no essential singularity at $q=0$ as functions of $q$: $\Gamma_1(1,2)$ has two cusps.}

Via the period map any section of $\lambda$ on $\mathcal A_g$ (resp. $\mathcal A_g^+$) can be pulled back to a section of $\lambda$ on $\mathcal M_g$ (resp. $\mathcal M_g^+$); the period map is holomorphic, so the pullback of a holomorphic section is holomorphic. For $g=2$ the pullback is bijective, for $g=3$ it is injective but not surjective \citep{Ichikawa95}, for $g\geq4$ it is neither. In other words, any section of the Hodge bundle on $\mathcal M_g$ or $\mathcal M_g^+$ extends to a Siegel modular form if $g=2$; if $g=3$, then it may not extend, but the extension is unique if it exists; and for $g\geq4$ an extension need not exist and need not be unique. In any case a meromorphic Siegel modular form of level $\Gamma_g$ (resp. $\Gamma_g(1,2)$) and weight $d$ always describes some well-defined meromorphic section of $\lambda^d$ on $\mathcal M_g$ (resp. $\mathcal M_g^+$) if $g\geq2$ or $\mathcal M_{1,1}=\mathcal A_1$ (resp. $\mathcal M_{1,1}^+=\mathcal A_1^+$) if $g=1$; a section described by a holomorphic modular form is holomorphic.

If $f$ is a meromorphic Siegel modular form of weight $d$, then, in accordance with the general notation of section~\ref{sec:orbifolds:bundles}, we denote the corresponding meromorphic section $s$ of $\lambda^d$ on $\mathcal M_g$, $\mathcal M_g^+$, $\mathcal M_{1,1}$ or $\mathcal M_{1,1}^+$ as
\begin{equation}
s = f\,dz^d.
\end{equation}

\textit{Remark.} 
A different parametrization of $\mathcal A_g^+$ is often used: $\mathcal A_g^+=\left(H_g\times \{\text{ev. ch.}\}\right)/\Gamma_g$, where $\{\text{ev. ch.}\}\subset(\mathbb Z/2)^{2g}\,=\{0,1\}^{2g}$ is the set of all even characteristics, a finite set known to consist of $2^{g-1}(2^g+1)$ elements. With this parametrization a section of the Hodge bundle on $\mathcal A_g^+$ corresponds to a function on $H_g\times\{\text{ev. ch}\}$, that is, to a set of functions $f[\delta]$ on $H_g$ labelled by even characteristics $\delta$. We do not use this parametrization in the present paper, but this parametrization is used by D'Hoker and Phong \citep{D'HoPho02}. It is easy to translate between the 2 parametrizations: the translation from $H_g\times\{\text{ev. ch}\}$ to $H_g$ is just the substitution $\delta=\begin{bmatrix}00...0\\00...0\end{bmatrix}$, and the translation in the inverse direction is described in \citep[section 2.7]{CDvG08}.

\section{The derivation}\label{sec:derivation}

Here we implement the plan of section~\ref{sec:plan}. The subsections of this section are numbered in the same way as the points of the plan.

\subsection{Line bundles}\label{sec:derivation:bundles}
It is well known from classical algebraic geometry (see e.g.\ \citep[section I.2]{ACGH1}) that a smooth genus 3 Riemann surface is either hyperelliptic or canonical. Hyperelliptic Riemann surfaces form a codimension 1 subspace $\mathcal H_3\subset \mathcal M_3$, so ``most'' genus 3 Riemann surfaces are canonical. We denote by $\mathcal M_{3,nh}\subset \mathcal M_3$ the moduli space of canonical genus 3 Riemann surfaces and $\mathcal M_{3,nh}^+\subset \mathcal M_3^+$ the moduli space of canonical genus 3 Riemann surfaces with an even spin structure.

It is explained in \citep{Witten13h} that the super Mumford form
\begin{equation}
\psi_3\in H^0\left(\mathcal S_3,\omega_{\mathcal S_3}\otimes b^{-5}\right),
\end{equation}
where $\omega_{\mathcal S_3}$ is the canonical line bundle on $\mathcal S_3$ (=the bundle of holomorphic Berezinian volume forms) and $b$ is the superanalogue of the Hodge bundle. One constructs an isomorphism $b\simeq\pi^*\lambda$ over $\mathcal M_{3,nh}^+$, where $\pi$ is the superperiod map, in the same way as for genus $2$ in \citep[the end of section 3.1.1]{Witten13h} or \citep[proposition 4.6]{FKP24}. This allows one to define\begin{equation}
\pi_*\psi_3\in H^0\left(\mathcal M_{3,nh}^+,\omega_{\mathcal M_{3,nh}^+}\otimes\lambda^{-5}\right)
\end{equation}
via fibrewise integration. All this is done in the same way as in \citep{Witten13h} or \citep{FKP24} for genus $2$, the only essential difference is that $\pi$ is not everywhere holomorphic for genus $3$, this is why we end up on $\mathcal M_{3,nh}^+$ and not on the whole $\mathcal M_{3}^+$.

\subsection{The ratio}\label{sec:derivation:ratio}

The Mumford form
\begin{equation}
\phi_3\in H^0\left(\mathcal M_3,\omega_{\mathcal M_3}\otimes\lambda^{-13}\right)
\end{equation}
is a trivialization of $\omega_{\mathcal M_3}\otimes\lambda^{-13}$ (a holomorphic section with no zeros), see \citep[section 2.1]{Witten13h}.

By pulling $\phi_3$ back from $\mathcal M_3$ to $\mathcal M_3^+$ we get an element of $H^0(\mathcal M_3^+,\omega_{\mathcal M_3^+}\otimes\lambda^{-13})$ that we also denote $\phi_3$, abusing notation. As $\phi_3$ has no poles and no zeros on $\mathcal M_3^+$, we may consider the quotient $\dfrac{\pi_*\psi_3}{\phi_3}$; it is be a holomorphic section of $\left(\omega_{\mathcal M_{3,nh}^+}\otimes\lambda^{-5}\right)\otimes\left(\omega_{\mathcal M_{3,nh}^+}\otimes\lambda^{-13}\right)^{-1} = \lambda^8$:
\begin{equation}
\dfrac{\pi_*\psi_3}{\phi_3}\in H^0\left(\mathcal M_{3,nh}^+,\lambda^8\right).
\end{equation}

This is convenient, because the canonical line bundle drops out.

\subsection{Determinantal representations}\label{sec:derivation:determinantal}

A Riemann surface of genus 3 is canonical $\Longleftrightarrow$ it is isomorphic to the zero set in $\mathbb P^2$ of a ternary quartic, i.e.\ a degree 4 homogeneous polynomial in 3 variables \citep[section I.2]{ACGH1}.

Let $V$ be the complex vector space of such polynomials; $\dim_{\mathbb C}V=15$. Those polynomials that define smooth curves in $\mathbb P^2$ form an open subset $V_0\subset V$; the complement to $V_0$ is a hypersurface, the zero set of the discriminant polynomial on $V$, see appendix~\ref{app:invariants:discriminant}.

Let $f(x)=\sum a_Ix^I=a_{400}x_0^4 + a_{310}x_0^3x_1 + a_{211}x_0^2x_1x_2 + ...$ be a ternary quartic from $V_0$. It is a classical fact that any such $f$ can be represented as the determinant of a symmetric $4\times4$ matrix $A(x)$ such that each entry $A_{kl}(x)$ of $A(x)$ is a linear form $A_{kl}(x)=x_0A_{0kl} + x_1A_{1kl}+x_2A_{2kl}$, $A_{ikl}\in\mathbb C$ \citep[proposition 4.2]{Beauville99}. The complex vector space $W$ of such matrices $A$ has dimension $30$. 2 groups act on $W$. The group $GL_3(\mathbb C)=GL_3$ acts by linear changes of variables $x_0,x_1,x_2$: $(Ag)(x)=A(gx)$ for $g\in GL_3$; and $GL_4(\mathbb C)=GL_4$ acts by conjugation: $h\in GL_4$ maps $A$ to $hAh^T$. (In other words, if we denote $E=\mathbb C^3$ and $F=\mathbb C^4$ the standard representation of $GL_3$ and $GL_4$ respectively, then $W=E^\vee\otimes Sym^2F^\vee$, where $\vee$ means the dual representation.) A matrix $k\,\mathrm{Id}_{4\times4}\in GL_4$ acts on $W$ in the same way as $k^2\,\mathrm{Id}_{3\times3}\in GL_3$ (here $k\in\mathbb C\setminus\{0\}$ and $\mathrm{Id}$ means the identity matrix), so we have defined an action on $W$ of the quotient group $G'=(GL_3\times GL_4) / \{k^{-2}\,\mathrm{Id}_{3\times3},k\,\mathrm{Id}_{4\times4}|k\in\mathbb C\setminus\{0\}\}$.

The action of $G'$ does not change the projective quartic curve $\det A(x)=0$. It is again a classical fact that there is a natural 1-to-1 correspondence between $G'$-orbits over a smooth quartic $f$ and even spin structures on the Riemann surface $C_f$ defined by the equation $f=0$ in $\mathbb P^2$, see \citep[lemma 6.3]{GroHar04} or \citep[proposition 4.2]{Beauville99} or \citep[theorem 4.1.3 and section 4.1.3]{Dolgachev}. From this one can deduce that $\mathcal M_{3,nh}^+  =W_0/G'$, where $W_0=\det^{-1}V_0$.

Analogously, $\mathcal M_{3,nh}=V_0/G$, where $G=GL_3 / \{k\,\mathrm{Id}_{3\times3}|k\in\mathbb C,k^4=1\}$ \citep[proposition 9.1]{CFvdG19}.\footnote{In \citep{CFvdG19} the authors twist the action of $GL_3$ by $\det^{-1}$; this is done in order to make the stabilizer of a generic quartic isomorphic to the automorphism group of a generic genus 3 Riemann surface, i.e.\ trivial. We do not twist the action but consider instead the quotient group $G=GL_3/\sqrt[4]1$; this is completely equivalent, because $\sqrt[4]1$ is the stabilizer of a generic quartic under the usual (i.e.\ not twisted) action of $GL_3$.} The 2 equivalences fit into the commutative diagram
\begin{equation}\label{eq:commutative-diagram-moduli}
\begin{tikzcd}
W_0/G'\ar[r,"\sim"]\ar[d,"\det"']&\mathcal M_{3,nh}^+\ar[d,"c"']\\
V_0/G\ar[r,"\sim"]&\mathcal M_{3,nh}
\end{tikzcd},
\end{equation}
where the right arrow means forgetting the spin structure.

\subsection{Sections of the Hodge bundle as invariants}\label{sec:derivation:hodge-as-invariants}

Now we want to describe sections of tensor powers of the Hodge line bundle $\lambda$ in terms of our parametrization, as in section~\ref{sec:orbifolds}. So we need a trivialization of the pullback of $\lambda$ to the space $W_0$ of parameters. If $F\in V_0$ and $C$ is the corresponding Riemann surface, then the three holomorphic 1-forms
\begin{equation}\label{eq:eta-b}
\eta_b = \operatorname{res}_Cx_b\dfrac{\frac12\epsilon_{ijk}x_idx_j\wedge dx_k}{F(x)}
\end{equation}
($b=0,1,2$) form a basis of the space of holomorphic 1-forms on $C$, see \citep[section 3.2]{LRZ08} for details. Here $\operatorname{res}$ is the Poincar\'e residue; in the part of $\mathbb P^2$ where $x_0\ne0$ we can set $x_0=1$ and use $x_1,x_2$ as coordinates; in these coordinates
\begin{equation}
\eta_b = \dfrac{x_b\,dx_2}{\frac{\partial f}{\partial x_1}(x_1,x_2)} = -\dfrac{x_b\,dx_1}{\frac{\partial f}{\partial x_2}(x_1,x_2)},
\end{equation}
where $f(x_1,x_2)=F(1,x_1,x_2)$ and $x_b=1$ when $b=0$. (We consider a non-singular Riemann surface: this means that at any $(x_1,x_2)\in\mathbb C^2$ satisfying $f(x_1,x_2)=0$ one has $\frac{\partial f}{\partial x_1}(x_1,x_2)\ne 0$ or  $\frac{\partial f}{\partial x_2}(x_1,x_2)\ne 0$, so at least 1 of the 2 expressions for $\eta_b$ is well-defined.) We choose 
\begin{equation}
\eta=\eta_0\wedge\eta_1\wedge\eta_2
\end{equation}
as our trivialization of $\lambda$ on $W_0$ (so that $\eta$ actually comes from $V_0$ as a pullback).

According to the general recipe of section~\ref{sec:orbifolds}, now we have to find out how the action of $GL_3\times GL_4$ affects $\eta$. Let $A\in W_0$, $F=\det A$, $k\in\mathbb C\setminus\{0\}$ and $g=k\,\mathrm{Id}_{3\times3}\in GL_3$. $(Ag)(x)=A(gx)=A(kx)=kA(x)$, so $(Fg)(x)=k^4F(x)$. Now it follows from the definition~\eqref{eq:eta-b} that $g$ acts on each $\eta_b$ as multiplication by $k^{-4}$, so it acts on $\eta$ as multiplication by $k^{-12}$. It follows easily from Hilbert's Nullstellensatz, as in \citep[proposition 3.2.1]{LRZ08}, that $SL_3\times SL_4$ acts trivially on $\eta$; so we have described the action of $G'$ on $\eta$ completely. According to section~\ref{sec:orbifolds}, this means that meromorphic sections of $\lambda^d$ on $\mathcal M_{3,nh}^+$ correspond bijectively to $(SL_3\times SL_4)$-invariant rational functions $\Phi$ on $W_0$ (equivalently, on $W$) such that $\Phi(kA)=k^{12d}\Phi(A)$, that is, to rational invariants of nets of quaternary quadrics of degree $12d$ (see appendix~\ref{app:invariants}).

\subsection{Regularity on \texorpdfstring{$\mathcal M_{3,nh}^+$}{\$\string\\mathcal M\_\{3,nh\}\string^+\$}}\label{sec:derivation:regularity}

By section~\ref{sec:orbifolds}, holomorphic sections of $\lambda^d$ correspond to rational invariants of degree $12d$ that are regular on $W_0$, that is, to those invariants that can be represented in the form $\frac{p(A)}{q(A)}$, where $p$ and $q$ are polynomials in $A_{ikl}$ and $q$ has no zeros on $W_0$.

By Salmon's theorem~\eqref{eq:salmons-theorem}, $\operatorname{discr}(\det(A)) = I(A)^2J(A)$ up to a constant factor, where $I$ and $J$ are certain polynomial invariants of degree $30$ and $48$ respectively; so $W\setminus W_0 = \{A\in W| I(A)=0$ or $J(A)=0\}$. One can check that $I$ and $J$ are irreducible polynomials, see appendix~\ref{app:invariants:irreducibility}. Now it follows from Hilbert's Nullstellensatz that any homogeneous rational invariant on $W$ regular on $W_0$ has the form $PI^aJ^b$, where $P$ is a polynomial invariant and $a,b$ are some integers (possibly zero or negative).

So it follows from sections~\ref{sec:derivation:ratio} and~\ref{sec:derivation:hodge-as-invariants} that 
\begin{equation}\label{eq:Xi3-PIaJb}
\dfrac{\pi_*\psi_3}{\phi_3}=PI^aJ^b\eta^8
\end{equation}
for some integers $a,b$ and some homogeneous polynomial invariant $P$. The degree of $PI^aJ^b$ must be $12\cdot8=96$, so $P$ is of degree $96-30a-48b$. $P$ is a polynomial, so its degree must be non-negative.

\subsection{Behaviour at infinity}\label{sec:derivation:infinity}

To get further, we consider the behaviour of $\dfrac{\pi_*\psi_3}{\phi_3}$ at infinity. First we recall some facts about compactifications of $\mathcal M_g, \mathcal M_g^+$ and $\mathcal S_g$. We use
\begin{itemize}
\item the Deligne-Mumford compactification $\bar{\mathcal M}_g$;
\item the compactification $\bar{\mathcal M}_g^+$ constructed by Cornalba (\citep{Cornalba89}, see also \citep{CCC04}) and, in another but equivalent way, by Jarvis (\citep{Jarvis95, Jarvis98, Jarvis99}, see also \citep{JKV99}), a review can be found in \citep{Sertoz17}; and
\item Deligne's compactification $\bar{\mathcal S}_g$ \citep{Deligne87, FKP20, MooZho19}.
\end{itemize}

We shall consider non-separating degenerations of Riemann surfaces of genus $g=3$
 (we only need genus 3, but the following holds for any $g\geq 2$). The closure of the collection of Riemann surfaces of arithmetic genus $g$ with exactly 1 singular point, a non-separating node, forms a divisor $D_0\subset\bar{\mathcal M}_g$, see e.g.\ \citep[section XII.10]{ACGH2}. Spin structures on singular Riemann surfaces of this kind are classified into 2 types: Ramond (R) and Neveu-Schwarz (NS). Accordingly, the preimage of $D_0$ in $\bar{\mathcal M}_g^+$ consists of 2 irreducible components $D_{0,R}$ and $D_{0,NS}$, see \citep[sections 4, 5]{Witten13h} or \citep[section 7]{Cornalba89} or \citep[section 3.2.2]{Jarvis99}. A particular superstructure was constructed on $D_{0,R}$ and $D_{0,NS}$, making them into divisors $\Delta_{0,R}$ and $\Delta_{0,NS}$ in $\bar{\mathcal S}_g$ \citep{FKP20}.

In the rest of this subsection we prove the following three statements:
\begin{enumerate}
\item $\pi_*\psi_3$ has an order $2$ pole at $D_{0,NS}$ and an order $1$ pole at $D_{0,R}$.
\item $\phi_3$ (pulled back to $\mathcal M_3^+$)  has an order $3$ pole at $D_{0,NS}$ and an order $2$ pole at $D_{0,R}$.
\item From the 2 previous statements it follows immediately that $\pi_*\psi_3/\phi_3$ has zeros of order $1$ at $D_{0,NS}$ and $D_{0,R}$. From this we shall deduce that $a=b=1$ in \eqref{eq:Xi3-PIaJb}.
\end{enumerate}

\subsubsection{Behaviour of \texorpdfstring{$\pi_*\psi_3$}{\$\string\\pi\_*\string\\psi\_3\$} near \texorpdfstring{$D_{0,NS}$}{\$D\_\{0,NS\}\$} and \texorpdfstring{$D_{0,R}$}{\$D\_\{0,R\}\$}} It is known that $\psi_g$ has an order $2$ pole at $\Delta_{0,NS}$ and an order $1$ pole at $\Delta_{0,R}$ for any $g\geq2$ \citep[theorem B]{FKP20}. Moreover, for $g=2$ or $3$ the fibrewise integration $\pi_*$ does not change the orders along non-separating boundary divisors, i.e.\ the order of $\pi_*\psi_g$ at $D_{0,NS}$ equals the order of $\psi_g$ at $\Delta_{0,NS}$, and the same holds for the R component. For $g=2$ this is proved in \citep[section 5]{Witten13h} via conformal field theory and in \citep{FKP24} via algebraic geometry, see proposition 7.9 in \citep{FKP24}.

The proof of \citep{FKP24} actually carries over to genus $3$, as we now explain. We need the genus $3$ analogue of proposition 6.2 and theorem 6.3(i) of \citep{FKP24}, and we only need the case of curves with just 1 singular point, a non-separating node. Inspecting the proofs in \citep{FKP24}, we find out that the part of proposition 6.2 devoted to this type of curves only depends on theorem 3.10(ii), which is valid for arbitrary genera.\footnote{The proof of proposition 6.2 in \citep{FKP24} also refers to a book by J. Fay. We note in passing that a mistake in Fay's formulas has been found \citep{Yamada80}, but this is not essential for the proof presented in \citep{FKP24}.} As for the proof of theorem 6.3(i) for this type of curves, the only thing that we need to change is the number of odd parameters: $\mathcal S_3$ has dimension $3g-3|2g-2\;=\;6|4$, so for genus $3$ we have not just $2$ odd parameters $\theta_1,\theta_2$ but 4 odd parameters $\theta_1,\theta_2,\theta_3,\theta_4$. So instead of $y=t+a\theta_1\theta_2$, $s(f)=t^2u=t^2(u_0+b\theta_1\theta_2)$ we have
\begin{equation}
\begin{array}{rl}
y&=t+\sum\limits_{1\leq i<j\leq 4}a_{ij}\theta_i\theta_j + a_{1234}\theta_1\theta_2\theta_3\theta_4,\\
s(f)&=t^2\left(u_0+\sum\limits_{1\leq i<j\leq 4}b_{ij}\theta_i\theta_j + b_{1234}\theta_1\theta_2\theta_3\theta_4\right),
\end{array}
\end{equation}
where $a_{ij}$ and $a_{1234}$ belong to $A_{bos}[t^{-1}]$, while $u_0$, $b_{ij}$ and $b_{1234}$ belong to $A_{bos}$. So by squaring the expression for $y$ we get
\begin{equation}
\begin{array}{ll}
y^2&=t^2 + 2t\left(\sum\limits_{1\leq i<j\leq 4}a_{ij}\theta_i\theta_j + a_{1234}\theta_1\theta_2\theta_3\theta_4\right) + (a_{12}a_{34}-a_{13}a_{24}+a_{14}a_{23})\theta_1\theta_2\theta_3\theta_4=\\
&=t^2\left(u_0+\sum\limits_{1\leq i<j\leq 4}b_{ij}\theta_i\theta_j + b_{1234}\theta_1\theta_2\theta_3\theta_4\right).
\end{array}
\end{equation}
Comparing these 2 expressions for $y^2$, we find by looking at the coefficient of $\theta_i\theta_j$ that actually $a_{ij}\in tA_{bos}$ for all $1\leq i<j\leq4$. Now we look at the coefficient of $\theta_1\theta_2\theta_3\theta_4$ and see that $2ta_{1234} + a_{12}a_{34}-a_{13}a_{24}+a_{14}a_{23}\in t^2A_{bos}$, so $2ta_{1234}\in t^2A_{bos}$, so $a_{1234}\in tA_{bos}$ too. Thus we see that $y\in tA$ for genus $3$ as well, just as in \citep{FKP24}, and this is enough to finish the proof.

So now we know that $\pi_*\psi_3$ has an order $2$ pole at $D_{0,NS}$ and an order $1$ pole at $D_{0,R}$.

\subsubsection{Behaviour of \texorpdfstring{$\phi_3$}{\$\string\\phi\_3\$} near \texorpdfstring{$D_{0,NS}$}{\$D\_\{0,NS\}\$} and \texorpdfstring{$D_{0,R}$}{\$D\_\{0,R\}\$}}

It is well known that $\phi_g$ has a pole of order 2 at the Deligne-Mumford boundary of $\mathcal M_g$ for any $g\geq2$ \citep[theorem XIII.7.15]{ACGH2}.
In other words, the divisor of $\phi_3$ on $\bar{\mathcal M}_3$ is
\begin{equation}
\operatorname{div}\phi_3=-2(D_0+D_1),
\end{equation}
where $D_1$ is the other boundary component of $\bar{\mathcal M}_3$ (corresponding to separating degenerations).
The projection $c:\bar{\mathcal M}_g^+\to\bar{\mathcal M}_g$ (forgetting the spin structure) is unramified (i.e.\ has ramification index $1$) at $D_{0,R}$ and has ramification index $2$ at $D_{0,NS}$ \citep[proposition 3.2.1]{Jarvis99}, so the pullback of the divisor $D_0$ via $c$ is
\begin{equation}\label{eq:pullback-of-D0}
c^*D_0=2D_{0,NS} + D_{0,R}.
\end{equation}
So the divisor of $\phi_3$ as a section of $c^*\omega_{\mathcal{\bar M}_3}\otimes\lambda^{-13}$ on $\mathcal{\bar M}_3^+$ would be
\begin{equation}
c^*\operatorname{div}\phi_3=-4D_{0,NS}-2D_{0,R} + \left(\text{some terms supported over }D_1\subset\bar{\mathcal M}_3\right).
\end{equation}
This is not quite what we want, because we want to pull back $\phi_3$ as a volume form, that is, we want a section of $\omega_{\bar{\mathcal M}_3^+}\otimes\lambda^{-13}$ on $\bar{\mathcal M}_3^+$, see section~\ref{sec:derivation:ratio}.

The Riemann-Hurwitz formula says that to pass from $c^*\omega_{\bar{\mathcal M}_3}$ to $\omega_{\bar{\mathcal M}_3^+}$ we must add a correction, the ramification divisor $R_c$ of $c$: by definition,
\begin{equation}
R_c=\sum\limits_D(e_D-1)D,
\end{equation}
where $D$ runs over the divisors where $c$ is ramified and $e_D$ is the ramification index of $c$ at $D$. The divisor of the pullback of $\phi_3$ to $\bar{\mathcal M}_3^+$ as a volume form is thus
\begin{equation}
c^*\operatorname{div}\phi_3 + R_c.
\end{equation}

The map $c$ is a covering over $\mathcal M_3\subset \bar{\mathcal M}_3$, so it is unramified there; thus it follows, again from~\eqref{eq:pullback-of-D0}, that
\begin{equation}
R_c=(2-1)D_{0,NS} = D_{0,NS}
\end{equation}
modulo terms supported over $D_1$. Summing the 2 contributions, we learn that the divisor of the pullback of $\phi_3$ to $\bar{\mathcal M}_3^+$ is
\begin{equation}
-4D_{0,NS}-2D_{0,R}+D_{0,NS} = -3D_{0,NS}-2D_{0,R}
\end{equation}
modulo terms supported over $D_1$, that is, $\phi_3$ has an order $3$ pole at $D_{0,NS}$ and an order 2 pole at $D_{0,R}$.\footnote{Here is a simple illustration. Consider the $1$-form $\alpha=z^{-k}dz$ on $\mathbb C$. Its divisor $\operatorname{div}\alpha=-kO$, where $O$ is the point $z=0$. Consider the map $c:\mathbb C\to\mathbb C$, $c(z)=z^e$, $e\ne 0$. Then $R_c=(e-1)O$. The Riemann-Hurwitz formula says that the divisor of the pullback of $\alpha$ via $c$ is $c^*\operatorname{div}\alpha+R_c=(-ek+e-1)O$. This is indeed the case: $c(z)^kdc(z)=z^{-ek}\cdot ez^{e-1}dz=ez^{-ek+e-1}dz$.}

It follows that $\pi_*\psi_3/\phi_3$ has zeros of order $1$ at $D_{0,NS}$ and $D_{0,R}$.

\subsubsection{Translation into the language of invariant theory} We now know that $\pi_*\psi_3/\phi_3=PI^aJ^b\eta^8$ has zeros of order $1$ at $D_{0,NS}$ and $D_{0,R}$, and we want to know what this says about the corresponding invariant $PI^aJ^b$. To this end, we consider the commutative diagram
\begin{equation}
\begin{tikzcd}
W_n/G'\ar[r,"m"]\ar[d,"\det"']&\bar{\mathcal M}_3^+\ar[d,"c"]\\
V_n/G\ar[r,"m_V"]&\bar{\mathcal M}_3
\end{tikzcd}
\end{equation}
extending~\eqref{eq:commutative-diagram-moduli}; here $V_n\subset V$ is the union of $V_0$ and the set of quartics that define curves in $\mathbb P^2$ of the type we have considered above (exactly 1 singular point, a non-separating node), $W_n$ is the preimage of $V_n$ in $W$, and the horizontal arrows are the classifying maps. It is known that the complement to $V_n$ in $V$ has codimension 2, so singular curves in $V_n/G$ form a divisor $D_{\operatorname{discr}}$ defined by the equation $\operatorname{discr}=0$; and it is known that
\begin{equation}
{m_V}^*D_0=D_{\operatorname{discr}},
\end{equation}
for example, this can be deduced from \citep[proposition 9.2]{CFvdG19}. 

From Salmon's theorem~\eqref{eq:salmons-theorem} it follows immediately that
\begin{equation}
\mathrm{det}^* D_{\operatorname{discr}} = 2D_I+D_J.
\end{equation}
We have already mentioned that
\begin{equation*}
c^*D_0=2D_{0,NS}+D_{0,R},
\end{equation*}
see \eqref{eq:pullback-of-D0}. Inspecting the commutative diagram, we find that $c(m(D_I))=m_V(\det(D_I))=D_0$, and analogously $c(m(D_J))=D_0$; so $m(D_I)$ is either $D_{0,NS}$ or $D_{0,R}$, and the same is true of $m(D_J)$. Moreover, the composition $c\circ m=m_V\circ\det$ has ramification index $2$ at $D_I$ and $1$ at $D_J$ (we find this by going through the bottom left corner of the diagram); so necessarily $m(D_J)=D_{0,R}$ with no ramification, and there are 2 possibilities for $m(D_{I})$: either $m(D_I)=D_{0,NS}$ with no ramification or $m(D_I)=D_{0,R}$ with ramification index $2$.

From the fact that the vanishing order of $\pi_*\psi_3/\phi_3$ at $D_{0,R}$ is $1$ and from $m(D_J)=D_{0,R}$ with no ramification we deduce that $b=1$ in the formula~\eqref{eq:Xi3-PIaJb}. The vanishing order of $\pi_*\psi_3/\phi_3$ at $D_{0,NS}$ is also $1$, so the first possibility for $m(D_I)$ would imply that $a=1$ and the second one would imply $a=2$. In the second case we would have $\deg P=96-2\cdot 30-48 < 0$, which is impossible; so the first possibility is the one that holds: $a=b=1$ and $\deg P=96-30-48=18$.

\subsection{Invariants of degree 18}\label{sec:derivation:18} 

Now we want to compute the dimension of the vector space of degree 18 invariants. This is a standard problem of representation theory. As we have mentioned, $W=E^\vee\otimes S^2F^\vee$, where $E=\mathbb C^3$ and $F=\mathbb C^4$ are the standard representation of $GL_3$ and $GL_4$ respectively and $S$ means symmetric power. So polynomial functions on $W$ are elements of $S^*(E^\vee\otimes S^2F^\vee)^\vee\simeq S^*(E\otimes S^2F)$. Thus we want to compute $\dim S^{18}(E\otimes S^2F)^{SL_3\times SL_4}$.

$M=k\,\mathrm{Id}_{3\times3}\in GL_3$ acts on $W$ as multiplication by $k$ and $N=k\,\mathrm{Id}_{4\times4}\in GL_4$ as multiplication by $k^2$. So a degree $d$ polynomial on $W$ scales by the factor of $k^d=(\det M)^{d/3}$ under the action of $M$ and by $k^{2d}=(\det N)^{d/2}$ under the action of $N$. In our case $d=18$, so, in other words, we want the dimension of the subrepresentation $\det_{GL_3}^6\boxtimes\det_{GL_4}^9$ of $GL_3\times GL_4$ in $W$.

We now describe the standard algorithm to find the multiplicities of irreducible subrepresentations in a given complex representation of a reductive algebraic group, specializing to the case of the group $GL_3\times GL_4$. 

Let $V$ be a complex representation of $GL_3\times GL_4$. A non-zero vector $v\in V$ is called a weight vector of weight $w=(a_1,a_2,a_3,b_1,b_2,b_3,b_4)\in\mathbb Z^7$ if $Dv=\prod\limits_{i=1}^3t_i^{a_i}\prod\limits_{j=1}^4u_j^{b_j}v$ for any $D=\left(\mathrm{diag}(t_1,t_2,t_3),\mathrm{diag}(u_1,u_2,u_3,u_4)\right)\in GL_3\times GL_4$. The character of $V$ is
\begin{equation}
\mathrm{ch}(V)=\sum\limits_wm_w\prod\limits_{i=1}^3t_i^{a_i}\prod\limits_{j=1}^4u_j^{b_j},
\end{equation} a polynomial in variables $t_1,...,u_4$ with integer coefficients; here $m_w$ is the dimension of the subspace of all weight $w$ vectors in $V$. $m_w\ne 0$ for at most $\dim V$ weights $w$; these weights are called the weights of $V$, and $m_w$ is called the multiplicity of $w$ in $V$.

For example, $E\otimes S^2F$ has a basis of 30 weight vectors $e_i\otimes f_jf_k$ ($1\leq i\leq3$, $1\leq j\leq k\leq4$), where $e_1,e_2,e_3$ and $f_1,...,f_4$ are the standard bases of $E$ and $F$ respectively. So $\mathrm{ch}(E\otimes S^2F)=(t_1+t_2+t_3)\left(u_1^2+u_2^2+u_3^2+u_4^2+u_1u_2+u_1u_3+u_1u_4+u_2u_3+u_2u_4+u_3u_4\right)$: there are $3\cdot10$ monomials here, one for each of the 30 basis vectors. Analogously, $W$ has a basis of weight vectors that consists of degree 18 monomials in the 30 basis vectors of $E\otimes S^2F$; the weight of such a monomial is the sum of the weights of the basis vectors of $E\otimes S^2F$ that occur in it. For example, $(e_2\otimes f_1^2)^{17}(e_3\otimes f_3f_4)$ has weight $17\cdot(0,1,0,2,0,0,0)+(0,0,1,0,0,1,1) = (0,17,1,34,0,1,1)$. Thus
\begin{equation}
\mathrm{ch}(W)=\sum\limits_{m}\prod\limits_{i=1}^3t_i^{a_i(m)}\prod\limits_{j=1}^4u_j^{b_j(m)},
\end{equation}
where the sum is over degree 18 monomials $m$ in the 30 basis vectors of $E\otimes S^2F$ and $w(m)=(a_1(m),...,b_4(m))$ is the weight of the monomial.

We order weights lexicographically: $w$ dominates $w'$ if $w_1\geq w'_1$, or ($w_1=w'_1$ and $w_2\geq w'_2$), or ($w_1=w'_1$ and $w_2=w'_2$ and $w_3\geq w'_3$), \&c. Any irreducible representation of $GL_3\times GL_4$ has a unique highest weight (that is, a weight that dominates any other weight of the representation), and an irreducible representation is determined uniquely up to an isomorphism by its highest weight. The highest weight of a representation is always dominant, that is, $w_1\geq w_2\geq ...\geq w_7$, and any dominant weight is the highest weight of an irreducible representation. If $R_w$ has highest weight $w$, then its character is the product of the corresponding Schur polynomials, that is,
\begin{equation}
\mathrm{ch}(R_w)=\sum\limits_T\prod\limits_{i=1}^3 t_i^{\#(i\text{ in }T)}
\cdot
\sum\limits_{U}\prod\limits_{j=1}^4 u_j^{\#(j\text{ in }U)},
\end{equation}
where the first sum is over semi-standard Young tableaux of shape $(a_1,a_2,a_3)$ (that is, at most 3 rows and the $i$'th row consists of $a_i$ boxes) filled with integers from the set $\{1,2,3\}$; $\#(i$ in $T)$ is the number of occurrences of the integer $i$ is the tableau $T$; the second sum is analogous.

A standard algorithm to find the multiplicities of irreducible subrepresentations of $W$ is as follows:

\begin{enumerate}

\item Start with the character $c=\mathrm{ch}(W)$ computed above, and set $n_w=0$ for all $w\in\mathbb Z^7$.

\item Find a maximal weight $w=(a_1,...,b_4)$ such that the coefficient of $\prod\limits_{i=1}^3t_i^{a_i}\prod\limits_{j=1}^4u_j^{b_j}$ is non-zero in $c$.

\item Increase $n_w$ by 1 and replace $c$ with $c-\mathrm{ch}(R_w)$.

\item If $c\ne 0$, then go to step $2$. If we have reached $c=0$, then $n_{(6,6,6,9,9,9,9)}$ is the dimension we seek.

\end{enumerate}

This algorithm leads to bulky calculations, so we used a computer to run it. The result is $n_{(6,6,6,9,9,9,9)}=3$, that is, the space of degree $18$ invariants is 3-dimensional.

In the literature we found three linearly independent degree $18$ invariants $\Lambda^3$, $I_3\Lambda$ and $Q'$, see \citep[section 5]{Gizatullin07} and appendix~\ref{app:invariants}.

Our result on the dimension then implies that

\begin{equation}\label{eq:Xi3-invariants:derivation}
\dfrac{\pi_*\psi_3}{\phi_3}=(k_1\Lambda^3 + k_2 I_3\Lambda + k_3Q')IJ\eta^8,
\end{equation}
where only $k_1,k_2,k_3\in \mathbb C$ remain to be determined.

\subsection{Correspondence between invariants and modular forms}\label{sec:derivation:modular-forms}

We have seen in section~\ref{sec:derivation:hodge-as-invariants} that meromorphic sections of $\lambda^d$ on $\mathcal M_{3,nh}^+$ correspond bijectively to rational invariants of nets of quadrics of degree $12d$. On the other hand, it is explained in section~\ref{sec:orbifolds:hodge} that meromorphic sections of $\lambda^d$ on $\mathcal A_3^+$ correspond bijectively to meromorphic Siegel modular forms of genus $3$, level $\Gamma_3(1,2)$ and weight $d$. The period map $\mathcal M_3^+\to \mathcal A_3^+$ pulls back sections of $\lambda^d$ on $\mathcal A_3^+$ to sections of $\lambda^d$ on $\mathcal M_3^+$; so the period map induces a linear map from meromorphic Siegel modular forms to rational invariants. This map is an injection, because the image of $\mathcal M_{3,nh}^+$ is dense in $\mathcal A_3^+$. By construction, a modular form $\alpha$ of weight $d$ maps to an invariant $C$ of degree $12d$ if and only if 
\begin{equation}
\alpha dz^d = C\eta^d
\end{equation}
as sections of $\lambda^d$ on $\mathcal M_{3,nh}^+$. In the end of this subsection we shall see that the invariant $PIJ$ describing $\dfrac{\pi_*\psi_3}{\phi_3}$ is in the image of this map.

\begin{enumerate}
\item First we consider the case $d=0$, i.e.\ we compare rational (=meromorphic) functions on $\mathcal M_3^+$ and $\mathcal A_3^+$. The period map is birational (it restricts to an embedding $\mathcal M_{3,nh}^+\to \mathcal A_3^+$ with dense image, cf.\ \citep{GuiMun95})
, so it induces an isomorphism of the spaces of rational functions; thus the map from meromorphic modular forms to rational invariants is an isomorphism in this case.

There is a holomorphic function $A:H_3\to W$ such that $A$ maps Jacobians of smooth Riemann surfaces to $W_0$ in such a way that the quartic $\det(A(\tau))$ with the even spin structure induced by this determinantal representation has, for some choice of a symplectic basis of the 1st homology, period matrix $\tau$ and theta characteristic $\begin{bmatrix}000\\000\end{bmatrix}$. Such a map is given in \citep[corollary 5.3]{DFS17}; it is not holomorphic on the whole $H_3$, only meromorphic, but it is easy to make it holomorphic by multiplying the matrix of \citep{DFS17} by some theta constants. The details and explicit formulas are given in appendix~\ref{app:map}; now we only want from $A$ the properties that we have just mentioned, and the precise form of $A$ is not important.

It follows that if $C$ is a rational invariant of degree $0$, then $C\circ A$ is the corresponding meromorphic Siegel modular form of weight $0$.

\item Now we consider the case $d=4$: we conjecture that the modular form $\theta_{00}^8$ (where $\theta_{00}=\theta\begin{bmatrix}000\\000\end{bmatrix}(\tau,0)$ is the theta constant) corresponds to the invariant $J$, up to a constant factor.  Here goes the argument:

\begin{enumerate}
\item By an old result of Klein, up to a constant factor
\begin{equation}
\chi_{18}dz^{18} = \operatorname{discr}^2\eta^{18}
\end{equation}
as meromorphic sections of $\lambda^{18}$ on $\mathcal M_{3,nh}$, where $\chi_{18}=\prod\limits_m\theta_m$ is the product of the 36 theta constants with even characteristics $m$; see \citep[proposition 4.1.2]{LRZ08}. Pulling this back to $\mathcal M_{3,nh}^+$ and applying Salmon's theorem~\eqref{eq:salmons-theorem}, we get
\begin{equation}
\chi_{18}dz^{18} = I^4J^2\eta^{18}
\end{equation}
on $\mathcal M_{3,nh}^+$, up to a constant factor.
\item We conjecture that the following relation holds for any $\tau\in H_3$:
\begin{equation}\label{eq:J5-over-I8}
\dfrac{\theta_{00}^{72}}{\chi_{18}^2}(\tau) = \dfrac{J^5}{I^8}(A(\tau)).
\end{equation}

Note that the modular form in the left-hand side has weight $\frac12\cdot 72-18\cdot 2=0$ and the invariant in the right-hand side has degree $48\cdot 5-30\cdot 8=0$.

This conjecture is supported by computer experiments. Namely, numerical calculations show that the relation~\eqref{eq:J5-over-I8} holds at many particular values of $\tau$. 
 

\item From the 2 previous equations it follows that $I^8J^4\eta^{36}\cdot\dfrac{J^5}{I^8} = \chi_{18}^2 dz^{36}\cdot\dfrac{\theta_{00}^{72}}{\chi_{18}^2}$, that is, $J^9\eta^{36}=\theta_{00}^{72}dz^{36}$, hence $J\eta^4=\theta_{00}^8dz^4$ (all equalities up to a constant factor).
\end{enumerate}

\item From points 1 and 2 it follows that
\begin{equation}
\dfrac{\pi_*\psi_3}{\phi_3} = PIJ\eta^8 = \frac{PI}{J}\cdot J^2\eta^8 = \Xi^{(3)}dz^8
\end{equation}
with
\begin{equation}
\Xi^{(3)}(\tau) =\frac{PI}{J}(A(\tau))\theta_{00}^{16}(\tau).
\end{equation}

Here $P=k_1\Lambda^3+k_2I_3\Lambda+k_3Q'$, and all ingredients are known but the three complex constants $k_1,k_2,k_3$. We have made use of the fact that $\dfrac{PI}{J}$ is of degree 0: $18+30=48$.

\end{enumerate}

\subsection{Factorization}\label{sec:derivation:factorization}
 
This subsection is devoted to formulating a conjecture on what the values of the three parameters in~\eqref{eq:Xi3-invariants:derivation} should be and providing evidence for this conjecture.

Namely, we want to go back to section~\ref{sec:intro:ansatz} and impose the ``factorization condition'', as it was done in \citep{CDvG08}:
\begin{equation}\label{eq:factorization-naive}
\Xi^{(3)}\begin{pmatrix}\tau_{11}&0&0\\0&\tau_{22}&\tau_{23}\\0&\tau_{23}&\tau_{33}\end{pmatrix}
=\Xi^{(1)}(\tau_{11})\Xi^{(2)}\begin{pmatrix}\tau_{22}&\tau_{23}\\\tau_{23}&\tau_{33}\end{pmatrix}.
\end{equation}
However, in this form the factorization condition does not make sense for our $\Xi^{(3)}$. Let us denote $H_1\times H_2\subset H_3$ the set of block-diagonal $3\times3$ matrices as in~\eqref{eq:factorization-naive}. $\Xi^{(3)}$ develops a pole along the divisor $\Theta'$ in $H_3$, where $\tau\in\Theta'$ iff $\theta_m(\tau)=0$ for some even characteristic $m\ne\begin{bmatrix}000\\000\end{bmatrix}$. This $\Theta'$ contains $H_1\times H_2$, so $\Xi^{(3)}$ has no limit at $\tau\in H_1\times H_2$ and~\eqref{eq:factorization-naive} does not make sense.

\subsubsection{A conjectural factorization condition}\label{sec:derivation:factorization:condition}

We may still try to compute the restriction of $\Xi^{(3)}$ to $H_1\times H_2$ ``by l'H\^opital's rule''. For any holomorphic function $f$ on the Siegel upper half-space $H_3$ one can compute the vanishing order of $f$ at $H_1\times H_2\subset H_3$ as the smallest integer $n$ such that for some $i$ ($0\leq i\leq n$) the value $\dfrac{\partial^n}{\partial\tau_{12}^{n-i}\partial\tau_{13}^i}f(\tau)\ne 0$ at some $\tau\in H_1\times H_2$. By the definition of theta constants, the restriction of $\theta_{00}(\tau)=\theta\!\begin{bmatrix}000\\000\end{bmatrix}\!(\tau)$ to $H_1\times H_2$ equals
$\theta\!\begin{bmatrix}0\\0\end{bmatrix}\!(\tau_{11})\;\,
\theta\!\begin{bmatrix}00\\00\end{bmatrix}\!\begin{pmatrix}\tau_{22}&\tau_{23}\\\tau_{23}&\tau_{33}\end{pmatrix}.$ In particular, $\theta_{00}$ does not vanish at a generic point of $H_1\times H_2$ (i.e.\ the vanishing order of $\theta_{00}$ at $H_1\times H_2$ is 0).

So we make the following conjecture:
\begin{enumerate}
\item The vanishing order
$n$ of $(PI)(A(\tau))$ at $H_1\times H_2$ is equal to the vanishing order of $J(A(\tau))$.

\item For any $i$ such that $0\leq i\leq n$ we have
\begin{equation}\label{eq:factorization}
\dfrac{D_{n,i}(PI)(A(\tau))}{D_{n,i}J(A(\tau))}\;\,
\theta^{16}\!\begin{bmatrix}0\\0\end{bmatrix}\!(\tau_{11})\;\,
\theta^{16}\!\begin{bmatrix}00\\00\end{bmatrix}\!
\begin{pmatrix}\tau_{22}&\tau_{23}\\\tau_{23}&\tau_{33}\end{pmatrix}
=\Xi^{(1)}(\tau_{11})\Xi^{(2)}\begin{pmatrix}\tau_{22}&\tau_{23}\\\tau_{23}&\tau_{33}\end{pmatrix},
\end{equation}
where 
\begin{equation}
D_{n,i}=\dfrac{\partial^n}{\partial\tau_{12}^i\partial{\tau_{13}^{n-i}}}\Bigg|_{\tau=\begin{pmatrix}\tau_{11}&0&0\\0&\tau_{22}&\tau_{23}\\0&\tau_{23}&\tau_{33}\end{pmatrix}} (0\leq i\leq n).
\end{equation}
\end{enumerate}

Strictly speaking, l'H\^opital's rule is not applicable in this situation, because $\Xi^{(3)}$ has no limit at $\tau\in H_1\times H_2$; but we make this conjecture nevertheless. The strategy of justifying the applicability of l'H\^opital's rule to our situation that we have in mind is as follows. The period map $\mathcal M_3\to \mathcal A_3$ contracts the divisor $D_1\subset\mathcal M_3$ corresponding to separating degenerations: the image of $D_1$ is the space $\mathcal A_1\times \mathcal A_2\subset \mathcal A_3$, which is not a divisor, because its codimension is 2. So it is natural to blow up $\mathcal A_1\times \mathcal A_2\subset \mathcal A_3$ if we want to study the restriction of $\dfrac{\pi_*\psi_3}{\phi_3}$ to $D_1$. The limits above are taken along paths issuing from $\mathcal A_1\times \mathcal A_2$ in normal directions, and normal directions are the points of the exceptional divisor of the blow-up. In this way we hope to deduce the factorization condition from \citep[theorem C]{FKP20} or a similar statement.

\subsubsection{Computer verification of the conjecture and the implied values of the three parameters} \label{sec:derivation:factorization:verification}

With a computer we obtained numerically the following results. The computer did not estimate error terms rigorously, so, strictly speaking, these results are only informal observations and not something obtained via a computer-assisted proof:

\begin{enumerate}
\item Point 1 of the conjecture of section~\ref{sec:derivation:factorization:condition} seems to hold for any degree 18 invariant $P$; the vanishing order $n=60$. To be more precise, we observed that for any degree 18 invariant $P\ne 0$ there exists an $i$ such that $D_{60,i}(PI)(A(\tau))$ does not vanish at some $\tau\in H_1\times H_2$, while for all $n<60$
$D_{n,i}(PI)(A(\tau))$ vanishes at all $\tau\in H_1\times H_2$ that we have tested (this has presently been checked for $i=0$ and $i=n=60$ only).

The same holds with $J(A(\tau))$ instead of $(PI)(A(\tau))$.

\item There exists just one degree 18 invariant $P$
such that~\eqref{eq:factorization} holds (this has also been checked for $i=0$ and $i=n=60$ only). With respect to the basis $\Lambda^3$, $I_3\Lambda$, $Q'$, this unique invariant $P$ has coordinates $(k_1,k_2,k_3)$ given by~\eqref{eq:3-parameters}.
\end{enumerate}

The very existence of such an invariant $P$ gives some evidence that our conjectured factorization condition should hold. If it does hold, then the values of the three coefficients (the components of $P$ with respect to the basis $\Lambda^3, I_3\Lambda, Q'$) follow from it, at least numerically.

\subsubsection{Remark: recovering the superstring measure for genus 1 and 2}\label{sec:derivation:factorization:recovering}

We find it noteworthy that we do not actually need to know explicit expressions for $\Xi^{(1)}$ and $\Xi^{(2)}$ to determine the coefficients \eqref{eq:3-parameters} up to proportionality from the factorization condition of section~\ref{sec:derivation:factorization:condition}. On the contrary, when applied to formula \eqref{eq:Xi3-modular:intro}, the factorization condition determines not only $P$ but $\Xi^{(1)}$ and $\Xi^{(2)}$ as well, up to constant factors. 

More precisely, we mean the following. Suppose that the equation
\begin{equation}\label{eq:factorization-recovering}
\dfrac{D_{60,i}\Big(\!\!\left(\kappa_1\Lambda^3+\kappa_2I_3\Lambda+\kappa_3Q'\right)I\Big)(A(\tau))}{D_{60,i}J(A(\tau))}\;\,
\theta^{16}\!\begin{bmatrix}0\\0\end{bmatrix}\!(\tau_{11})\;\,
\theta^{16}\!\begin{bmatrix}00\\00\end{bmatrix}\!
\begin{pmatrix}\tau_{22}&\tau_{23}\\\tau_{23}&\tau_{33}\end{pmatrix}
=\xi_1(\tau_{11})\,\xi_2\!\begin{pmatrix}\tau_{22}&\tau_{23}\\\tau_{23}&\tau_{33}\end{pmatrix}
\end{equation}
holds for some functions $\xi_1:H_1\to\mathbb C$ and $\xi_2:H_2\to\mathbb C$, for some complex numbers $\kappa_1,\kappa_2,\kappa_3$, for all block-diagonal matrices $\tau\in H_1\times H_2\subset H_3$ and both for $i=0$ and $i=60$. Numerical calculations suggest that in this case the triple $(\kappa_1,\kappa_2,\kappa_3)$ must be proportional to the triple \eqref{eq:3-parameters} and the functions $\xi_1,\xi_2$ must coincide with $\Xi^{(1)},\Xi^{(2)}$ up to constant factors.
We believe that a rigorous analytic proof of this assertion can be obtained by analyzing the left-hand side of \eqref{eq:factorization-recovering}.

This means that the genus 1 and the genus 2 superstring measure can be recovered, up to constant factors, from our (almost) rigorously derived formula \eqref{eq:Xi3-modular:intro} for $\Xi^{(3)}$ and the conjectural factorization condition exposed in section~\ref{sec:derivation:factorization:condition}; that is, from \eqref{eq:Xi3-modular:intro} and \eqref{eq:factorization-recovering}. We do not rely here on any known properties of the superstring measures for genus 1 and 2; in particular, we obtain a formula for the genus 2 superstring measure (up to a constant factor) independently of D'Hoker and Phong's results.

In the rest of this section we describe our numerical computer calculations that have led us to the conclusions we have just discussed. We did not estimate error terms rigorously during the calculations, so, strictly speaking, all results in this section are informal observations (they will hopefully be turned into rigorous proofs in the future). Below we shall use our formula \eqref{eq:Xi3-modular:intro} for the superstring measure, the factorization condition \eqref{eq:factorization-recovering} and the formula \eqref{eq:Xi1} for $\Xi^{(1)}$ to produce the triple \eqref{eq:3-parameters} and D'Hoker and Phong's modular form $\Xi^{(2)}$ (see \eqref{eq:Xi2}), up to constant factors. 
In the same manner one can do more: one can obtain \eqref{eq:3-parameters}, $\Xi^{(1)}$ and $\Xi^{(2)}$ (up to constant factors) from \eqref{eq:Xi3-modular:intro} and \eqref{eq:factorization-recovering} only, using neither \eqref{eq:Xi1} nor \eqref{eq:Xi2}. This is possible at the expense of making the calculations a little more complicated: below we only have to solve a system of linear equations, but we should have to solve a quadratic system if we avoided using \eqref{eq:Xi1}.

\begin{enumerate}
\item By evaluating numerically at several particular block-diagonal matrices $\tau\in H_1\times H_2\subset H_3$ we observed that
\begin{equation}\label{eq:factorization-independence-of-tau11}
\dfrac{D_{60,i}\Big(\!\!\left(\kappa_1\Lambda^3+\kappa_2I_3\Lambda+\kappa_3Q'\right)I\Big)(A(\tau))}{D_{60,i}J(A(\tau))}\;\,
\cdot\dfrac{\theta^{16}\!\begin{bmatrix}0\\0\end{bmatrix}\!(\tau_{11})}{\Xi^{(1)}(\tau_{11})}
\end{equation}
does not depend on $\tau_{11}\in H_1$ when the variables $\begin{pmatrix}\tau_{22}&\tau_{23}\\\tau_{23}&\tau_{33}\end{pmatrix}\in H_2$, $\kappa:=\begin{pmatrix}\kappa_1\\\kappa_2\\\kappa_3\end{pmatrix}\in\left\{\begin{pmatrix}1\\0\\0\end{pmatrix},\begin{pmatrix}0\\1\\0\end{pmatrix},\begin{pmatrix}0\\0\\1\end{pmatrix}\right\}$ and $i\in\{0,60\}$ are kept fixed. As \eqref{eq:factorization-independence-of-tau11} is linear in $\kappa$, this independence holds equivalently for all $\kappa\in\mathbb C^3$.

The said independence implies that if \eqref{eq:factorization-recovering} holds for some $\xi_1,\xi_2$, for some $\kappa\in\mathbb C^3$ and for some $i\in\{0,60\}$ (and for all $\tau\in H_1\times H_2$), then necessarily $\xi_1=u\Xi^{(1)}$ for some $u\in\mathbb C$.

Let us denote \eqref{eq:factorization-independence-of-tau11} by $L_i(\tau_{22},\tau_{23},\tau_{33})\kappa$; we shall assume that $L_i(\tau_{22},\tau_{23},\tau_{33})$ is a row vector here, which is multiplied by the column vector $\kappa$.

\item Then we searched for $\kappa\in\mathbb C^3$ such that
\begin{equation}\label{eq:factorization-linear-equation}
L_0(\tau_{22},\tau_{23},\tau_{33})\kappa
=L_{60}(\tau_{22},\tau_{23},\tau_{33})\kappa.
\end{equation}
This is a linear equation in $\kappa\in\mathbb C^3$ when $\begin{pmatrix}\tau_{22}&\tau_{23}\\\tau_{23}&\tau_{33}\end{pmatrix}\in H_2$ is fixed.

We picked 3 different period matrices $\tau^{(1)},\tau^{(2)},\tau^{(3)}\in H_2$ and thus obtained a system
\begin{equation}\label{eq:factorization-system}
M\left(\tau^{(1)},\tau^{(2)},\tau^{(3)}\right)\kappa=0
\end{equation}
of 3 linear equations in $\kappa\in\mathbb C^3$; here $M\left(\tau^{(1)},\tau^{(2)},\tau^{(3)}\right)$ is the complex $3\times3$ matrix whose $i$'th row is $L_0\left(\tau^{(i)}_{22},\tau^{(i)}_{23},\tau^{(i)}_{33}\right) - L_{60}\left(\tau^{(i)}_{22},\tau^{(i)}_{23},\tau^{(i)}_{33}\right)$.

Then we evaluated the characteristic polynomial $p:=\det\left(M\left(\tau^{(1)},\tau^{(2)},\tau^{(3)}\right)-\lambda\,\mathrm{Id}_{3\times3}\right)$ and observed that the constant term of $p$ was $0$ and the term linear in $\lambda$ was non-zero. This should mean that $0$ is an eigenvalue of $M\left(\tau^{(1)},\tau^{(2)},\tau^{(3)}\right)$ and the multiplicity of this eigenvalue is $1$, so the system \eqref{eq:factorization-system} has just 1 solution up to proportionality. We computed this unique solution and observed that it was proportional to \eqref{eq:3-parameters}.

We repeated this for several triples $\left(\tau^{(1)},\tau^{(2)},\tau^{(3)}\right)$ and got the same result.

\item Finally, we evaluated
\begin{equation} 
L(\tau):=L_0(\tau_{22},\tau_{23},\tau_{33})k\;\theta^{16}\!\begin{bmatrix}00\\00\end{bmatrix}(\tau)
\end{equation}
with $k$ given by \eqref{eq:3-parameters} at several particular $\tau=\begin{pmatrix}\tau_{22}&\tau_{23}\\\tau_{23}&\tau_{33}\end{pmatrix}\in H_2$.
We also evaluated $\Xi^{(2)}(\tau)$ directly via the formula \eqref{eq:Xi2} at the same matrices $\tau\in H_2$. In all cases we observed that $L(\tau)=\Xi^{(2)}(\tau)$.

Thus our numerical computations suggest that if \eqref{eq:factorization-recovering} holds for some $\xi_1,\xi_2$ and for some $\kappa\in\mathbb C^3$ (and for all $\tau\in H_1\times H_2$, $i\in\{0,60\}$), then $\xi_1=u\Xi^{(1)}$, $\xi_2=v\Xi^{(2)}$ and $\kappa=uvk$ with $k$ given by \eqref{eq:3-parameters} for some $u,v\in\mathbb C$.

\end{enumerate}

We regard this observation both as an additional check of our formulas \eqref{eq:Xi3-invariants:intro}--\eqref{eq:Xi3-modular:intro} and a heuristic argument in favour of the validity of the factorization condition of section~\ref{sec:derivation:factorization:condition}.

\section{The sum over spin structures, and the 0-point function}\label{sec:vacuum}

In section~\ref{sec:vacuum:interior} we prove that the superstring measure $\pi_*\psi_3$ vanishes when summed over spin structures, i.e.\ over the fibres of the projection $c:\mathcal M_3^+\to\mathcal M_3$. The analogous statement for genus 2 was proved by D'Hoker and Phong \citep[section 6.3]{D'HoPho01-4}.
For our proof we only need the formula \eqref{eq:Xi3-invariants:intro} with three unknown parameters (which has been derived rigorously): the vanishing holds for any triple $(k_1,k_2,k_3)$, so we do not have to rely on our conjecture \eqref{eq:3-parameters} about the values of these three parameters.

In section~\ref{sec:vacuum:boundary} we review the relation between summation over spin structures and the vacuum amplitude (a.k.a.\ 0-point function) of type II superstring theory on $\mathbb R^{10}$. It is expected that the genus $g$ contribution to the 0-point function should vanish for any $g$ \citep{Martinec86}. We explain that choosing a holomorphic projection $\pi:\mathcal S_{3,nh}\to\mathcal M_{3,nh}^+$ (where $\mathcal S_{3,nh}\subset \mathcal S_3$ is the open sub-orbifold over $\mathcal M_{3,nh}^+$) splits the genus 3 0-point function into 2 summands, which are sometimes called the ``contribution from the interior of the moduli space'', i.e.\ from $\mathcal S_{3,nh}$, and the ``contribution from the boundary'', i.e.\ from a ``complement'' to $\mathcal S_{3,nh}$ inside $\bar{\mathcal S}_3$. This splitting into 2 summands depends on the choice of $\pi$. The contribution from the interior is the integral of the ``modulus squared'' of the sum of $\pi_*\psi_3$ over spin structures. As the sum over spin structures vanishes pointwise for our choice of $\pi$ (i.e.\ even before taking the modulus squared and integrating it), the contribution from the interior is 0. In order to confirm the vanishing of the genus 3 0-point function it remains to prove that the contribution from the boundary vanishes too. The analysis of the boundary contribution is left for future research.

\subsection{The sum of the genus 3 superstring measure \texorpdfstring{$\pi_*\psi_3$}{\$\string\\pi\_*\string\\psi\_3\$} over spin structures vanishes}\label{sec:vacuum:interior}

Let $c_*$ denote the sum over even spin structures, i.e.\ over the fibres of the $36$-sheeted covering map $c:\mathcal M_{3,nh}^+\to\mathcal M_{3,nh}$. In this section we prove that
\begin{equation}
c_*\pi_*\psi_3=0.
\end{equation}

We shall be actually computing $c_*\left(\dfrac{\pi_*\psi_3}{\phi_3}\right)$: $\phi_3$ has no dependence on spin structures, so $c_*\left(\dfrac{\pi_*\psi_3}{\phi_3}\right)=\dfrac{c_*\pi_*\psi_3}{\phi_3}$, and
$\phi_3$ has no zeros and no poles in the interior of the moduli space, so
$c_*\left(\dfrac{\pi_*\psi_3}{\phi_3}\right)=0$ is equivalent to $c_*\pi_*\psi_3=0$.

We have 2 expressions for $\dfrac{\pi_*\psi_3}{\phi_3}$: via modular forms \eqref{eq:Xi3-modular:intro} and via invariants \eqref{eq:Xi3-invariants:intro}. Consequently, there are 2 equivalent ways to compute $c_*\pi_*\psi_3$. Namely, the following 3 statements are equivalent:

\begin{itemize}
\item $c_*\pi_*\psi_3$ vanishes at a point $C\in\mathcal M_{3,nh}$.

\item $\sum\limits_\delta\Xi^{(3)}[\delta](\tau)=0$, where $\tau\in H_3$ is a period matrix for $C$.

Here the sum is over the 36 even characteristics $\delta$, $\Xi^{(3)}[0]:=\Xi^{(3)}$ and $\Xi^{(3)}[\delta]$ for $\delta\ne0$ is obtained from $\Xi^{(3)}$ as described in \citep[section 2.7]{CDvG08}, see also the remark at the end of section~\ref{sec:orbifolds:hodge} for more details on this.

\item $\left(c_*(PIJ)\right)(f)
:=\sum\limits_{k=1}^{36}(PIJ)(A[k])=0$, where $P$ is the degree 18 invariant of nets making the formula \eqref{eq:Xi3-invariants:intro} true, $f$ is a ternary quartic such that the Riemann surface in $\mathbb P^2$ defined by the equation $f=0$ is isomorphic to $C$ and $A[k]\in W$ ($k=1,...,36$) are nets of quadrics such that the $(GL_3\times GL_4)$-orbits of $A[k]$ are pairwise distinct and $\det A[k]=f$ for all $k$.
\end{itemize}

We shall prove the last of these equivalent statements. Even more generally, let $P$ be any degree 18 invariant of nets of quadrics; we shall now prove that $c_*(PIJ)$ vanishes identically.

\begin{enumerate}
\item $PIJ$ has degree $\deg P +\deg I+\deg J=18+30+48=96=12\cdot8$, so it corresponds to a certain section $s:=PIJ\eta^8\in H^0\left(\mathcal M_{3,nh}^+,\lambda^8\right)$, see  section~\ref{sec:derivation:hodge-as-invariants}. By integrating $s$ along the fibres of $c$ we get the section $c_*s = \left(c_*(PIJ)\right)\eta^8\in H^0\left(\mathcal M_{3,nh},\lambda^8\right)$. This implies that $c_*(PIJ)$ is a polynomial invariant of ternary quartics of degree $3\cdot8=24$ (as a polynomial in the 15 coefficients of a ternary quartic); see \citep[proposition 3.3.1]{LRZ08}, this is analogous to the statement of section~\ref{sec:derivation:hodge-as-invariants}, but for $\mathcal M_{3,nh}$ instead of $\mathcal M_{3,nh}^+$.

\item $c_*(PIJ)$ is defined (has a finite limit) at any singular ternary quartic $f\in V$. Indeed, if $\operatorname{discr}f=0$, then for any $A\in W$ with $\det A=f$ one has either $I(A)=0$ or $J(A)=0$ by Salmon's theorem \eqref{eq:salmons-theorem}; as $PIJ$ contains both $I$ and $J$ as factors, one has $\left(c_*(PIJ)\right)(f)=0$.

Thus $c_*(PIJ)$ vanishes on the set $\{\mathrm{discr}=0\}\subset V$.

\item Now it follows from Hilbert's Nullstellensatz and from the irreducibility of $\mathrm{discr}$ that the polynomial $c_*(PIJ)$ is divisible by $\mathrm{discr}$. But $\deg\mathrm{discr}=27>24=\deg (c_*(PIJ))$, so $c_*(PIJ)=0$. This finishes the proof.
\end{enumerate}

\emph{Remark.}
The argument actually shows that $c_*R=0$ for any homogeneous polynomial invariant $R$ of nets of quadrics such that $R$ is divisible by $IJ$ and the degree of $R$ is divisible by $12$ and  $<108=27\cdot4$. But, as one can prove, these conditions imply that either $R=k\Lambda IJ$ with $k\in\mathbb C$ or $R=PIJ$ with $P$ of degree $18$.

\subsection{The 0-point function of type II superstring theory}\label{sec:vacuum:boundary}
Here we comment on the relation between $c_*\pi_*\psi_3$ and the genus 3 0-point function of type II superstring theory. This section is a review of known results.

\subsubsection{The 0-point function as an integral of the ``modulus squared'' of the super Mumford form}

The genus $g$ 0-point function of type II superstring theory on flat $\mathbb R^{10}$ is known to be equal to a certain integral of the ``modulus squared'' of $\psi_g$ over the supermoduli space. We need some notation to state this more precisely:

\begin{itemize}
\item For a complex (super)manifold $M$ we denote the complex conjugate (super)manifold by $M^{op}$. This means that the underlying topological space of $M^{op}$ is equal to the underlying topological space $|M|$ of $M$, and the structure sheaf $\mathcal O_{M^{op}}=\mathbb C^{op}\otimes_{\mathbb C}\mathcal O_M$, where $\mathbb C^{op}$ is the ring $\mathbb C$ with the usual right $\mathbb C$-algebra structure and the left $\mathbb C$-algebra structure $z\cdot w:=\bar zw$ ($z\in\mathbb C, w\in\mathbb C^{op}$, $\bar z$ is the complex number conjugate to $z$). In the non-super case, $M^{op}$ is isomorphic to $M$ with its sheaf of holomorphic functions replaced by that of anti-holomorphic ones. We denote the $\mathbb C$-antilinear isomorphism $\mathcal O_M\to\mathcal O_{M^{op}}$ as $f\mapsto\tilde f:=1\otimes f$ for $f\in\mathcal O_M(U)$, $U\subset|M|$ an open subset (so that $\tilde{kf}=\bar k\tilde f$ for any $k\in\mathbb C$). Analogously, if $s$ is a section of a vector bundle $E$ on $M$, we denote by $\tilde s$ the corresponding section $1\otimes s$ of $E^{op}:=\mathcal O_{M^{op}}\otimes_{\mathcal O_M}E$, and if $\phi:M\to N$ is a morphism of (super)manifolds, then we denote by $\phi^{op}$ the corresponding morphism $M^{op}\to N^{op}$. This notation generalizes routinely to (super)orbifolds. We identify $\mathrm{Ber}(M^{op})$ with $\mathrm{Ber}(M)^{op}$ (where $\mathrm{Ber}$ denotes the line bundle of holomorphic Berezinian volume forms), etc.

\item $\Gamma\subset \mathcal S_g\times \mathcal S_g^{op}$ is a CS sub-superorbifold of maximal odd dimension such that its bosonic truncation $\Gamma_{bos}\subset \mathcal M_3^+\times \mathcal M_3^{+op}$ coincides with the \emph{quasi-diagonal} (or is sufficiently close to the quasi-diagonal) \citep[section 3.1.1]{Witten13h}, \citep[section 3.3.2]{Witten12p}. The notion of a CS sub-supermanifold is a generalization to the super case of the notion of a real submanifold in a complex manifold, see Witten's notes \citep{Witten12i} or \citep[secton 4.8]{DelMor99}. The quasi-diagonal consists of pairs $((\Sigma,L),(\Sigma,L'))\in\mathcal M_3^+\times{\mathcal M}_3^{+op}$, where $\Sigma$ is a Riemann surface and $L$, $L'$ are two even spin structures on $\Sigma$ ($L$ and $L'$ may coincide or not coincide). We need the quasi-diagonal and not just the diagonal because in type II superstring theory ``left'' and ``right'' spin structures are independent of each other. 

\item Let $p_{\mathcal S_g}:\mathcal S_g\times \mathcal S_g^{op}\to \mathcal S_g$ and $p_{\mathcal S_g^{op}}:\mathcal S_g\times \mathcal S_g^{op}\to \mathcal S_g^{op}$ be the canonical projections. There is a canonical isomorphism $p_{\mathcal S_g}^*\mathrm{Ber}(\mathcal S_g)\otimes p_{\mathcal S_g^{op}}^*\mathrm{Ber}(\mathcal S_g^{op})\simeq \mathrm{Ber}(\mathcal S_g\times \mathcal S_g^{op})$, so $p_{\mathcal S_g}^*\psi_g\otimes p_{\mathcal S_g^{op}}^*\tilde{\psi_g}$ can be considered as a holomorphic Berezinian volume form on $\mathcal S_g\times \mathcal S_g^{op}$ valued in the line bundle $p_{\mathcal S_g}^*b^5\otimes p_{\mathcal S_g^{op}}^*(b^{op})^5$; let us denote this holomorphic Berezinian volume form by  $p_{\mathcal S_g}^*\psi_g\wedge p_{\mathcal S_g^{op}}^*\tilde{\psi_g}$ (with $\wedge$ instead of $\otimes$). Applying to $p_{\mathcal S_g}^*\psi_g\wedge p_{\mathcal S_g^{op}}^*\tilde{\psi_g}$ the natural pairing $h^5:p_{\mathcal S_g}^*b^5\otimes p_{\mathcal S_g^{op}}^*(b^{op})^5\to\mathcal O_{\mathcal S_g\times \mathcal S_g^{op}}$ defined near the quasi-diagonal (see \citep[section 3.1.1]{Witten13h} or \citep[section 5.2]{FKP19} for details), we get a holomorphic Berezinian volume form (now valued in the trivial line bundle) that we denote by $|\psi_g|^2$; it is defined on a neighbourhood of the quasi-diagonal in $\mathcal S_g\times\mathcal S_g^{op}$.

\end{itemize}

The genus $g$ 0-point function of type II superstring theory equals
\begin{equation}\label{eq:vacuum}
\int\limits_\Gamma|\psi_g|^2
\end{equation}
\citep[section 3.1.1]{Witten13h}. Unfortunately, what we have just said is not enough for \eqref{eq:vacuum} to be well-defined. The reason is that $\mathcal S_g$ is not compact; this leads to 2 problems:

\begin{enumerate}
\item The integral \eqref{eq:vacuum} depends on the choice of $\Gamma$; the restrictions on $\Gamma$ stated above are not enough to fix this ambiguity.

\item It is not clear a priori whether the integral \eqref{eq:vacuum} is convergent. It actually turns out that the integral only converges conditionally and a regularization is needed. Cf.~\citep[section 2.4]{Witten13p}.\footnote{A problem with convergence is already manifest in bosonic string theory, making it inconsistent. It is believed that in superstring theory, as opposed to bosonic string theory, the integrals can be regularized and evaluated consistently to give finite values of amplitudes.}
\end{enumerate}

In section~\ref{sec:vacuum:boundary:compact} we consider a simplified situation: we suppose that the domain of integration is compact, so these 2 problems do not arise. We shall explain that in the compact case any holomorphic projection can be used to compute the integral --- if at least one holomorphic projection exists, but note that there are complex supermanifolds that do not admit any holomorphic projections to the bosonic truncation \citep[section 2.3.1]{Witten12i}, \citep{DonWit13}.

In section~\ref{sec:vacuum:boundary:non-compact} we shall return to our non-compact situation and explain that our result on the vanishing of the sum over spin structures implies the vanishing of the contribution from the interior of the moduli space to the genus 3 0-point function.

\subsubsection{Computing integrals by using holomorphic projections: the compact case}\label{sec:vacuum:boundary:compact}

Let us consider the following simplified situation. Let $S$ be a complex supermanifold, $M^+:=S_{bos}$ its bosonic truncation (obtained by setting all odd coordinates to zero), and suppose that $M^+$ is \emph{compact}. Let $c:M^+\to M$ a finite-sheeted (unramified) covering of compact complex manifolds. Let $\psi$ be a holomorphic Berezinian volume form on $S$. Let $\Gamma\subset S\times S^{op}$ be a CS sub-supermanifold of maximal odd dimension (i.e.\ $\Gamma$ has twice as many odd coordinates as $S$) such that $\Gamma_{bos}\subset M^+\times M^{+op}$ is the quasi-diagonal $\{(s,s')\in M^+\times M^{+op}\;|\;c(s)=c^{op}(s')\}$. (The reader will not lose really much by assuming that $M^+=M$, $c$ is the identity map and $\Gamma_{bos}\subset M\times M^{op}$ is the diagonal.) We want to compute the integral
\begin{equation}\label{eq:vacuum-compact}
\int\limits_{\Gamma}|\psi|^2,
\end{equation}
where $|\psi|^2$ is our shorthand for $p_S^*\psi\wedge p_{S^{op}}^*\tilde\psi$
(as above); this integral is well-defined, because $\Gamma$ is compact. Note that in general $\Gamma$ is not defined uniquely by what we have said: there can be many ways to extend the quasi-diagonal to a CS sub-supermanifold of maximal odd dimension. However, in the compact case this ambiguity does not affect the value of the integral \eqref{eq:vacuum-compact}.

Let $\pi:S\to M^+$ be a holomorphic projection (i.e.\ a morphism of complex supermanifolds such that $\pi i=\mathrm{id}$, where $i:M^+\to S$ is the canonical embedding, corresponding to setting all odd coordinates to zero). Then
\begin{equation}\label{eq:vacuum-compact-projection}
\int\limits_\Gamma|\psi|^2 = 
\int\limits_{M}\left|c_*\pi_*\psi\right|^2,
\end{equation}
where $M$ in the right-hand side is considered as a real manifold, $\pi_*$ is the integration along the fibres of $\pi$ and $\left|c_*\pi_*\psi\right|^2$ is the shorthand for the wedge product of the differential form $c_*\pi_*\psi$ on the real manifold $M$ and the differential form complex conjugate to it.

This statement may be considered as a kind of the Fubini theorem: it amounts to replacing the ``multiple'' integral \eqref{eq:vacuum-compact} with the following ``iterated'' integral.
Let $\pi_\Gamma:\Gamma\to\Gamma_{bos}$ be the restriction of the projection $\pi\times\pi^{op}:S\times S^{op}\to M^+\times M^{+op}$ to $\Gamma$. In this situation
\begin{equation}
\int\limits_{\Gamma}|\psi|^2
= \int\limits_{\Gamma_{bos}}{\pi_\Gamma}_*\left(|\psi|^2\right),
\end{equation}
see \citep[corollary 2.12]{AHP11}; this is where compactness (of $\Gamma_{bos}$) is essential. As both $\pi_{\Gamma}$ and $\psi$ have the form (something) $\times$ (its complex conjugate) --- namely, $\pi_{\Gamma}$ is the restriction of $\pi\times\pi^{op}$ and $|\psi|^2:=p_S^*\psi\wedge p_{S^{op}}^*\tilde\psi$ --- we have
\begin{equation}\label{eq:vacuum-compact-integrand}
{\pi_\Gamma}_*\left(|\psi|^2\right)
\;=\;
p_{M^+}^*\left({\pi}_*\psi\right)\;\wedge\; p_{M^{+op}}^*\tilde{\left({\pi}_*\psi\right)},
\end{equation}
where $p_{M^+}:\Gamma_{bos}\to M^+$ and $p_{M^{+op}}:\Gamma_{bos}\to M^{+op}$ are the projections.

Now we use that $\Gamma_{bos}$ is the quasi-diagonal, so the map $q:\Gamma_{bos}\to\Delta$ induced by $M^+\times M^{+op}\xrightarrow{c\times c^{op}}M\times M^{op}$ is a $(36\cdot36)$-sheeted covering; here $\Delta\subset M\times M^{op}$ is the diagonal (so $\Delta$ is isomorphic to $M$ considered as a real manifold). Thus integration over $\Gamma_{bos}$ is equivalent to summation over the fibres of $q$ followed by integration over $\Delta\simeq M$. By construction, $p_{M^+}^*({\pi}_*\psi)$ is constant along the fibres of $\Gamma_{bos}\hookrightarrow M^+\times M^{+op}\xrightarrow{\mathrm{id}\times c^{op}}M^+\times M^{op}$ and $p_{M^{+op}}^*\tilde{({\pi}_*\psi)}$ is constant along the fibres of $\Gamma_{bos}\hookrightarrow M^+\times M^{+op}\xrightarrow{c\times\mathrm{id}}M\times M^{+op}$, so by summing the integrand \eqref{eq:vacuum-compact-integrand} over the fibres of $q$ we get
\begin{equation}
\int\limits_{\Gamma_{bos}}p_{M^+}^*\left({\pi}_*\psi\right)\wedge p_{M^{+op}}^*\tilde{\left({\pi}_*\psi\right)}
= \int\limits_M\left(c_*\pi_*\psi\right)\wedge\bar{\left(c_*\pi_*\psi\right)}\\
=: \int\limits_M\left|c_*\pi_*\psi\right|^2,
\end{equation}
where $M$ is considered as a real manifold and the horizontal bar means the complex conjugate differential form.

In particular, the value of the integral \eqref{eq:vacuum-compact} does not depend on the choice of $\Gamma$ (satisfying the requirements given above) and the value of the right-hand side of \eqref{eq:vacuum-compact-projection} does not depend on the choice of a holomorphic projection $\pi$. We stress that compactness of $M^+$ is crucial here.

If $M^+$ is not compact, then the left-hand side of \eqref{eq:vacuum-compact-projection} may depend non-trivially on the choice of $\Gamma$ and the right-hand side on the choice of $\pi$, making the equation \eqref{eq:vacuum-compact-projection} invalid. Sometimes this dependence is referred to by saying that the contribution from the ``boundary'' of $M^+$ (i.e.\ the complement to $M^+$ inside some compactification of $M^+$) should be taken into account.

\subsubsection{Holomorphic projections in the non-compact case, and the boundary contribution to the 0-point function}\label{sec:vacuum:boundary:non-compact}

The discussion above implies that the pointwise vanishing of $c_*\pi_*\psi_3$ on $\mathcal M_{3,nh}$ would imply the vanishing of the genus 3 $0$-point function $\int\limits_\Gamma|\psi_3|^2$ if $\mathcal M_{3,nh}$ were compact. As $\mathcal M_{3,nh}$ is not compact, the equality $\int\limits_\Gamma|\psi_3|^2 = \int\limits_{\mathcal M_{3,nh}}\left|c_*\pi_*\psi_3\right|^2$ will only hold if $\Gamma$ and $\pi$ are compatible in some sense. Superstring theory is actually believed to fix a choice of the integration cycle $\Gamma$ (not completely, but in such a way that after a certain regularization procedure the integral in the left-hand side of \eqref{eq:vacuum} becomes well-defined); a great part of \citep{Witten12p} is devoted to sketching the construction of $\Gamma$ and the necessary regularization procedure. So the question now is whether our projection $\pi$ (induced by the superperiod map) is compatible with the particular class of $\Gamma$'s described in \citep{Witten12p}. In other words, the integral
\begin{equation}
\int\limits_{M_{3,nh}}\left|c_*\pi_*\psi_3\right|^2
=\int\limits_{M_{3,nh}}|0|^2
=0
\end{equation}
is the contribution from the ``interior'' $\mathcal S_{3,nh}$ into the genus 3 0-point function. To confirm that the genus 3 0-point function vanishes, it remains to prove that the ``boundary contribution'' vanishes, i.e.\ that $\pi$ is compatible with $\Gamma$. We leave the analysis of the boundary contribution for future research.

In the rest of this subsection we collect some references. A construction of the integration cycle $\Gamma$ and the necessary regularization procedure is sketched in \citep[right up to section 6.6]{Witten12p} and \citep{Witten13p}. For genus 2, the vanishing of the boundary contribution is discussed in \citep[section 3.2]{Witten13p} and also in \citep[section 19]{D'Hoker14}. The genus $2$ computation has been revisited by Felder, Kazhdan and Polishchuk (not published yet); they claim to have worked out all details about the integration cycle and the regularization procedure for genus 2, which allowed them to prove the vanishing of the genus 2 0-point function on the mathematical level of rigour, although they had to use not the projection $\pi$ coming from the superperiod map but another class of algebraic projections to compute the integral \citep{Kazhdan24}. There is an attempt to analyze the boundary contribution for genus 3 \citep{BettLin20}, but in any case it does not cover the hyperelliptic locus, which should also be considered as a part of the boundary when dealing with our projection $\pi$ since $\pi$ is not holomorphic at the hyperelliptic locus. $\pi$ is only a holomorphic projection $\mathcal S_{3,nh}\to\mathcal M_{3,nh}^+$; it is an open question whether a holomorphic projection $\mathcal S_3\to\mathcal M_3^+$ exists, see \citep{DonWit13}.

\section{Conclusions \& further directions}\label{sec:conclusion}

Here we summarize the results of this paper and indicate some questions for further research.

We have obtained the formula for $\dfrac{\pi_*\psi_3}{\phi_3}$ in 2 forms: in terms of invariant theory and in terms of modular forms.

\begin{enumerate}
\item The derivation of the invariant theory version of the formula is sufficiently rigorous. As for the translation into the language of modular forms, it remains to prove the relation~\eqref{eq:J5-over-I8} rigorously, now it is only observed to hold in numerical experiments.

\item Our formula contains three unknown parameters. We conjecture the values of the parameters, but further effort is needed to prove (or maybe disprove) this conjecture. Namely, to prove the conjecture, one should prove that
\begin{enumerate}
\item $\Xi^{(3)}$ should satisfy a factorization condition, that is, its restriction to $H_1\times H_2$ (in the appropriate sense) should coincide with $\Xi^{(1)}(\tau_{11})\Xi^{(2)}\begin{pmatrix}\tau_{22}&\tau_{23}\\\tau_{23}&\tau_{33}\end{pmatrix}$, and that
\item the regularized restriction procedure of section~\ref{sec:derivation:factorization} is a valid one.
\end{enumerate}
We hope that this can be deduced with the help of \citep[theorem C]{FKP20}, similarly to what is done for genus $2$ in \citep[section 7]{FKP24}.
\end{enumerate}
Furthermore, it would be interesting to check whether our $\Xi^{(3)}$ has some other properties expected from superstring theory:

\begin{enumerate}
\setcounter{enumi}{2}

\item By using our formula we can compute the vanishing orders of $\pi_*\psi_3$ at the 2 divisors in $\mathcal M_3^+$ lying over the hyperelliptic divisor in $\mathcal M_3$ to be $4$ and $-4$, thus re-deriving Witten's result \citep[appendix C.4]{Witten15}. This is going to be addressed in a future publication.

\item We have proved that the sum of $\pi_*\psi_3$ over spin structures vanishes. It remains to analyze the boundary behaviour of $\psi_3$ to check whether our formula for $\pi_*\psi_3$ implies the vanishing of the 0-point function in type II superstring theory.

\item Just like the 0-point function, 1-, 2- and 3-point functions are also expected to vanish \citep{Martinec86}, and for genus 1 and 2 they also vanish pointwise after summation over spin structures, even without integration over $\mathcal M_{1,1}$ and $\mathcal M_2$, see \citep[section 12.1]{D'HoPho02}. It may be possible to use our formula in order to check whether the analogous vanishing holds for genus 3.

\item Using their formulas for $\psi_2$, D'Hoker and Phong also computed some non-vanishing 2-loop amplitudes \citep{D'HoPho02}. It is interesting whether our formula for $\pi_*\psi_3$ can be applied to compute some non-vanishing 3-loop amplitudes, for example, 4-point functions. Low-energy expansions of some non-vanishing 3-loop amplitudes were computed in the pure spinor formalism in \citep{GomMar13}, but no corresponding computation in the RNS formalism has been carried out yet, and no non-vanishing 3-loop superstring amplitudes have been computed completely by now.

\item\label{sec:conclusion:ansatz}
We should like to clarify the relation between our formula for $\Xi^{(3)}$ and the Ansatz of \citep{CDvG08}. The latter Ansatz was used in a recent work \citep{GMS21} to make a proposal for the complete genus 3 4-point amplitude, and this proposal matches some partial results on 3-loop amplitudes form \citep{GomMar13}. Perhaps the ratio of our $\Xi^{(3)}$ to the Ansatz of \citep{CDvG08} is a modular function with some nice properties; such a relation might explain why both our expression for $\Xi^{(3)}$ and the Ansatz of \citep{CDvG08} imply results matching other known results.

\item The idea about algebraic parametrizations that we have used to derive a formula for $\pi_*\psi_3$ can in principle be applied to genera $g\geq3$ to derive formulas for $\psi_g$, or maybe for some components thereof, maybe after the restriction to some part of the (super)moduli space, --- as long as one can find an amenable algebraic parametrization of (some part of) the (super)moduli space.

\item\label{sec:conclusion:coefficients}
What is the nature of the coefficients $k_1=87026940\cdot1197218880$ and $k_3=-87026940$ given in~\eqref{eq:3-parameters}? They are special in that they do not contain large prime factors. Actually they can be written as multinomial coefficients
:
\begin{equation}
k_1=\binom{28}{8,7,5,5,1,1,1}=\frac{28!}{8!\cdot7!\cdot5!\cdot5!}
\end{equation}
and
\begin{equation}
k_3=-\binom{28}{20,6,2}=-\frac{28!}{20!\cdot6!\cdot2}.
\end{equation}
It is not clear whether this is a coincidence or there is a reason behind: we only obtained these coefficients numerically. But in any case it is noteworthy that these coefficients are integers, a priori they could be arbitrary complex numbers.

Of course, the values of the coefficients depend on the normalization of the invariants $\Lambda$ and $Q'$; the natural normalization that we have chosen is described at the end of appendix~\ref{app:invariants:invariants}.

\item The algebra of modular forms of genus $3$ and level $\Gamma_{3}(2,4)$ is generated by second order theta constants with a single algebraic relation between them \citep[section 3]{Runge95}. Perhaps it would be illuminating to re-write our formula in terms of second order theta constants.

\item Can the formulas for $\Xi^{(1)}, \Xi^{(2)}$ and $\Xi^{(3)}$ be naturally written in a uniform way?

\item $\pi_*\psi_3$ does not describe $\psi_3$ completely, only its term of degree 4 in odd variables with respect to local coordinates compatible with the projection $\pi$; see \citep[eq. (3.9)]{Witten13h}. D'Hoker and Phong \citep{D'HoPho02} found a formula not only for $\pi_*\psi_2$ (the term of degree 2 in odd variables with respect to local coordinates compatible with $\pi$) but also for $\psi_2\Big|_{\mathcal M_2^+}$ (the degree 0 term; this component has an invariant meaning, independent of $\pi$). So, all in all, D'Hoker and Phong derived a complete formula for $\psi_2$. It would be interesting, and potentially useful for computing superstring amplitudes, to find formulas for other parts of $\psi_3$ (for the invariant degree 0 term, and also for the terms of degree 2 in odd variables with respect to some local coordinates).

\end{enumerate}

\appendix

\section{Theta functions}\label{app:theta}

Here we fix our notation and terminology for theta functions; our notation is similar to that of \citep{CDvG08} or \citep{DFS17}.

Let the genus $g\geq1$ be fixed. A characteristic is a vector $m=\begin{bmatrix}a_1&a_2&\dots a_g\\b_1&b_2&\dots b_g\end{bmatrix}$ such that each $a_i$ or $b_i$ is either $0$ or $1$; we consider $a_i$ and $b_i$ as integers modulo $2$. A characteristic $m$ is even (resp.\ odd) if its parity $a^Tb=\sum\limits_{i=1}^ga_ib_i=0$ (resp. $1$); here $a$ and $b$ are interpreted as column vectors and $M^T$ means the transpose of a matrix $M$.

The Siegel upper half-space $H_g\subset\mathbb C^{g\times g}$ is the set of symmetric $g\times g$ matrices with positive definite imaginary part.

For a fixed $\tau\in H_g$ the theta function with characteristic $m$ is the function of $z=(z_0,z_1,...,z_{g-1})\in\mathbb C^g$ given by
\begin{equation}
\theta_m(\tau,z)=\theta\begin{bmatrix}a_1&a_2&\dots a_g\\b_1&b_2&\dots b_g\end{bmatrix}(\tau,z)=\sum\limits_{n\in\mathbb Z^g}\exp2\pi i\left[\frac12\left(n+\frac a2\right)^T\!\!\tau\left(n+\frac a2\right) + \left(n+\frac a2\right)^T\!\!\left(z+\frac b2\right)\right].
\end{equation}

$\theta_m(\tau)=\theta_m(\tau,0)$ is called the theta constant with characteristic $m$ (it is zero if $m$ is odd).

For $g=3$ we use a shorthand notation: a characteristic $\begin{bmatrix}a_2&a_1&a_0\\b_2&b_1&b_0\end{bmatrix}$ is encoded by 2 decimal digits $4a_2+2a_1+a_0$ and $4b_2+2b_1+b_0$; e.g.\ $04$ means the characteristic $\begin{bmatrix}000\\100\end{bmatrix}$ and $\theta^{12}_{04}$ means the $12$'th power of the theta constant with this characteristic.

A set $\{m_1,m_2,m_3\}$ of three distinct characteristics is called syzygetic if the parity of $m_1+m_2+m_3$ equals the sum of the parities of $m_1$, $m_2$ and $m_3$, otherwise the triple is azygetic. A set $\{m_1,m_2,,...,m_k\}$ of $k\geq3$ distinct characteristics is called syzygetic (resp.\ azygetic) if every subset of three characteristics is syzygetic (resp.\ azygetic).

\section{Selected invariants of nets of quaternary quadrics}\label{app:invariants}

Here we give explicit formulas for the invariants of nets of quaternary quadrics that we use. This appendix is essentially an extraction from \citep{Gizatullin07}, except for the last part~\ref{app:invariants:irreducibility}.

\subsection{Invariants}\label{app:invariants:invariants}

We denote $E=\mathbb C^3$ and $F=\mathbb C^4$ the standard representations of $GL_3$ and $GL_4$ respectively and $W=E^\vee\otimes S^2F^\vee$, where $\vee$ is the dual representation. An element of $E^\vee$ is a linear function on $E$ and an element of $S^2F^\vee$ is a quadric on $F$, so an element of $W$ can be though of as a linear function on $E$ valued in quadrics on $F$. Quadrics on $F=\mathbb C^4$ are identified with symmetric $4\times 4$ matrices, so an element $A\in W$ can also be thought of as a symmetric matrix $A(x)=x_0A_0+x_2A_1+x_2A_2$ (where $x_0,x_1,x_2$ is the standard basis of $E^\vee$) or a triple of symmetric $4\times 4$ matrices $A_0,A_1,A_2$.

$W$ is naturally isomorphic to the space of linear maps from $E$ to $S^2F^\vee$, so $A\in W$ defines a vector subspace of the space $S^2F^\vee$ of quadrics (the image of the whole $E$). For a generic $A$ this subspace has dimension $3=\dim E$, thus such an $A$ defines a 2-dimensional projective subspace --- a net --- in $\mathbb PS^2F^\vee$. Therefore $W$ is sometimes called (with a slight abuse of language) the space of nets of quaternary quadrics.

A polynomial (resp.\ rational) invariant $F$ of nets of quaternary quadrics is a polynomial (resp.\ rational) $(SL_3\times SL_4)$-invariant function on $W$. Explicitly this means that
\begin{equation}
\begin{array}{ll}
F(A_0,A_1,A_2)&=F\left(\sum_iM_{0i}A_i, \sum_iM_{1i}A_i, \sum_iM_{2i}A_i\right)\\
&=F\left(N^TA_0N, N^TA_1N, N^TA_2N\right)
\end{array}
\end{equation}
for any $M\in SL_3$ and $N\in SL_4$

The union of orbits of nets of the special form \citep[example 2.7]{Gizatullin07}
\begin{equation}
A(x)=\begin{pmatrix}0&ax_0+bx_1+cx_2&ex_0+fx_1+gx_2&px_0+qx_1+rx_2\\
ax_0+bx_1+cx_2&0&x_2&x_1\\
ex_0+fx_1+gx_2&x_2&0&x_0\\
px_0+qx_1+rx_2&x_1&x_0&0
\end{pmatrix}
\end{equation}
is dense in $W$, so an invariant of nets is determined completely by its values on nets of this form.\footnote{The notation that Gizatullin uses in his example 2.7 is inconsistent with the rest of his paper; to make it consistent, one should change $f,g,h$ in the formulas of example 2.7 with $e,f,g$ respectively.} 
So we give the formulas for invariants either as polynomials in the entries of three general symmetric matrices $A_0,A_1,A_2$ or as polynomials in $a,b,c,e,f,g,p,q,r$.

An $SL_3$-invariant polynomial (or rational) function on $V:=S^4E^\vee$ is called an invariant of ternary quartics. For any invariant $F:V\to\mathbb C$ of ternary quartics its pullback $F\circ\det:W\to\mathbb C$ via $\det:W\to V$ is an invariant of nets of quaternary quadrics. The invariant $I_3$ (section~\ref{app:invariants:I3}) and the discriminant (section~\ref{app:invariants:discriminant}) are obtained in this way.

The invariants described below are normalized in such a way that they have, as polynomials in $a,b,c,e,f,g,p,q,r$, integer coefficients with no common multiple. This condition fixes the normalization constant up to a sign. 

\subsubsection{Degrees of invariants}

We say that a polynomial invariant $\Phi$ of nets of quadrics has degree $d$ if
\begin{equation}\label{eq:degree}
\Phi(kA)=k^d\Phi(A)
\end{equation}
for any $k\in\mathbb C$ and $A\in W$; in other words, if the function $\Phi$ is homogeneous of degree $d$. More generally, we say that a rational invariant $\Phi$ is of degree $d$ if \eqref{eq:degree} holds for all $k\in\mathbb C$ and all $A\in W$ such that $\Phi$ is defined at $A$. We do not apply the word ``degree'' to invariants that are not homogeneous.

The degree of any homogeneous polynomial invariant $\Phi$ is twice as big as the degree of the corresponding (necessarily homogeneous) polynomial in $a,b,c,e,f,g,p,q,r$ and 4 times as big as what Gizatullin calls ``order'' in \citep{Gizatullin07}.

Analogously, a homogeneous invariant $F$ of ternary quartics is of degree $d$ if $F(k^df)=k^dF(f)$ for all $k\in\mathbb C$ and all ternary quartics $f\in V$ such that $F$ is defined at $f$. The degree of the corresponding invariant of nets of quadrics $F\circ\det$ is $4d$, because $\det:W\to V$ has degree $4$.

\subsection{The Toeplitz invariant \texorpdfstring{$\Lambda$}{\$\string\\Lambda\$} of degree 6}\label{app:invariants:Lambda}
In the general situation
\begin{equation}
\Lambda(A)=\operatorname{Pf}\begin{pmatrix}0&-A_2&A_1\\A_2&0&-A_0\\-A_1&A_0&0\end{pmatrix},
\end{equation}
where $\operatorname{Pf}$ is the Pfaffian of a skew-symmetric $12\times12$ matrix \citep[section 2]{Gizatullin07}.

As a polynomial in $a,b,c,e,f,g,p,q,r$,

\begin{equation}
\begin{array}{ll}
\Lambda =& a(g^2-q^2)+f(p^2-c^2)+r(b^2-e^2)+\\
&+(bcg+bgp+egp+bpq)-(ceg+bcq+ceq+epq)
\end{array}
\end{equation}
 \citep[eq. (2.5)]{Gizatullin07}.

\subsection{The invariant \texorpdfstring{$I_3$}{\$I\_3\$} of degree 12}\label{app:invariants:I3}
\begin{equation}
I_3(A)
=\dfrac1{2^8\cdot3^2}\det
\begin{pmatrix}
\dfrac{\partial}{\partial x_0} & \dfrac{\partial}{\partial x_1} & \dfrac{\partial}{\partial x_2}\\[2ex]
\dfrac{\partial}{\partial y_0} & \dfrac{\partial}{\partial y_1} & \dfrac{\partial}{\partial y_2}\\[2ex]
\dfrac{\partial}{\partial z_0} & \dfrac{\partial}{\partial z_1} & \dfrac{\partial}{\partial z_2}\\
\end{pmatrix}^4
f(x)f(y)f(z),
\end{equation}
where $f(x)=\det A(x)$.

An explicit formula for $I_3$ as a polynomial in $a,b,c,e,f,g,p,q,r$ takes a whole page \citep[appendix 14.1]{Gizatullin07} (so we do not repeat it).

$I_3$ is (the pullback of) an invariant of ternary quartics. As an invariant of ternary quartics, $I_3$ has degree $3 = 12 / 4$.

\subsection{The invariant \texorpdfstring{$Q'$}{Q'} of degree 18}\label{app:invariants:Qprime}

\begin{itemize}
\item We denote by $A_i(u)$ the quadric $\sum\limits_{k,l=0}^3A_{ikl}u_ku_l$ in 4 variables $u_0,u_1,u_2,u_3$, here $A_{ikl}$ is the entry of $A_i$ in the $k$'th row and the $l$'th column.

\item We denote by $\widehat A_i$ the symmetric complex matrix with entries $\widehat A_{ikl}=\dfrac{1+\delta_{kl}}2\dfrac{\partial\Lambda}{\partial A_{ikl}}(A)$ ($0\leq i\leq 2$, $0\leq k\leq l\leq 3$); here $\Lambda$ is considered as a polynomial in the entries $A_{ikl}$ ($0\leq k\leq l\leq 3$) of the symmetric matrices $A_i$.

\item We denote by $J(A,u)$ the $3\times4$ matrix
$\dfrac{\partial A_i(u)}{\partial u_l}$
($0\leq i\leq 2$, $0\leq l\leq 3$) and by $J_k(A,u)$ the $3\times3$ matrix obtained from $J(A,u)$ by removing the $k$'th column ($0\leq k\leq 3$). We let 
$X_k(A,u)$ ($k=0,1,2,3$) be the polynomial
$(-1)^k\det J_k(A,u);$ this is a cubic polynomial in $u_0,u_1,u_2,u_3$ whose coefficients depend polynomially on $a,b,c,e,f,g,p,q,r$.

\item We denote $p_{kk}(A)=X_k(\widehat A,\frac{\partial}{\partial u})X_k(A,u)$, where $X_k(\widehat A,\frac{\partial}{\partial u})$ is the differential operator obtained by substituting $\frac{\partial}{\partial u_i}$ instead of $u_i$ into $X_k(\widehat A,u)$.

\item Finally,
\begin{equation}
Q'(A)=\dfrac1{2^7}\sum\limits_{k=0}^3p_{kk}(A).
\end{equation}

\end{itemize}

An explicit formula for $Q'$ as a polynomial in $a,b,c,e,f,g,p,q,r$ takes two pages, for the sake of completeness it is given in appendix~\ref{app:Qprime}.

\textit{Remark.} Gizatullin does not use the notation $Q'$. He focuses instead on another degree $18$ invariant $Q$ defined as some linear combination of $Q'$ and $\Lambda^3$ \citep[eq. (5.2)]{Gizatullin07}. We are not sure what exact linear combination he prefers, because he leaves an undetermined coefficient $c$ in his formula (5.2).

\subsection{The Salmon invariant \texorpdfstring{$I$}{I} of degree 30}
\begin{equation}
\begin{array}{ll}
I=&(be-af)(ar-cp)(gq-fr)\times\\
&\times(rb^2-fc^2+(g-q)bc)(fp^2-aq^2+(b-e)pq)(ag^2-re^2+(p-c)eg)
\end{array}
\end{equation}
\citep[theorem 7.2, proof]{Gizatullin07}.

\subsection{The tact invariant \texorpdfstring{$J$}{J} of degree 48}

Here we follow \citep[section 10]{Gizatullin07} and use Gizatullin's notation (which we explain here).

\begin{equation}
J=(afrd)^2J',
\end{equation}
where
\begin{equation}
d=\det\begin{pmatrix}a&b&c\\e&f&g\\p&q&r\end{pmatrix},
\end{equation}

$$J'=\dfrac{1}{16F^2}\det\begin{pmatrix}
4A & 3B & 2C & G & 0 & 0\\
0  & 4A & 3B & 2C & G & 0\\
0  & 0  & 4A & 3B & 2C & G\\
B & 2C & 3G & 4E & 0 & 0\\
0 & B & 2C & 3G & 4E & 0\\
0 & 0 & B & 2C & 3G & 4E\\
\end{pmatrix},$$
here $A,B,C,E,F,G$ are polynomials in $a,b,c,e,f,g,p,q,r$ that we shall define momentarily; $J'$ looks like a rational function, but in fact this rational function is a polynomial in $a,b,c,e,f,g,p,q,r$, i.e.\ $F^2$ divides the determinant.
\begin{equation}
A = ar^2 - cpr
\end{equation}
(of course, this is not the matrix $A\in W$ used in previous subsections: we have a small conflict of notation here),
\begin{equation}
B = cpq + bcr- cgp- cer + bpr + 2agr- 2aqr- c^2q,
\end{equation}

\begin{equation}
C = bcg - ceg + cfp + bgp + bcq +ceq,
\end{equation}

\begin{equation}
E = af^2 - bef,
\end{equation}

\begin{equation}
F=b^2r- c^2f +bcg- bcq,
\end{equation}

\begin{equation}
\begin{array}{ll}
G =& bcf + cef +beg- bfp- beq + 2afq- 2afg- b^2g-\\
&-bpq + ber- 2agq- 2afr + ag^2 + aq^2 - b^2r- c^2f.
\end{array}
\end{equation}

\subsection{The discriminant and Salmon's theorem}\label{app:invariants:discriminant}
Consider the space $V=S^4E^\vee$ of ternary quartics. The equation $f=0$ for $f\in V$ defines a singular quartic in $\mathbb P^2$ if and only if $f$ belongs to the zero set of a certain irreducible polynomial on $V$. This polynomial is called the discriminant of ternary quartics, and we denote it $\operatorname{discr}$; it is defined uniquely up to a constant factor by what we have just said. $\operatorname{discr}$ is a degree $27$ invariant of ternary quartics. See \citep[section 13.D]{GKZ}.

The discriminant of a ternary quartic $f$ can be represented as the determinant of a $15\times15$ matrix $M$ such that each matrix element occurring in the first 9 lines of $M$ depends linearly on the coefficients of $f$ and each matrix element from the 6 remaining lines of $M$ is a homogeneous cubic polynomial in the coefficients of $f$. This formula is attributed to Gordan; see \citep{Edge48}, or \citep[chapter 13, proposition 1.6]{GKZ} for a modern treatment.

\textit{Salmon's theorem.} The invariant of nets of quadrics corresponding to $\mathrm{discr}$ coincides with $I^2J$ up to a constant factor, where $I$ and $J$ are defined above. In other words, for any $A\in W$
\begin{equation}\label{eq:salmons-theorem}
\operatorname{discr}(\det A))=kI^2(A)J(A),
\end{equation}
where $k$ is a constant independent of $A$. See \citep[corollary 10.4]{Gizatullin07}. (The value of $k$ depends on the constant factor in the definition of $\mathrm{discr}$ that we have not specified.)

\subsection{Irreducibility of \texorpdfstring{$I$}{I} and \texorpdfstring{$J$}{J}}\label{app:invariants:irreducibility}

Here we prove that the Salmon invariant $I$ and the tact invariant $J$ are irreducible polynomials (or rather we indicate several theorems in the literature that combine into a proof). Cf.\ the discussion around \citep[theorem 7.5]{PSV10}. This irreducibility is crucial for our argument in sections \ref{sec:derivation:regularity} and \ref{sec:derivation:infinity}.

\begin{itemize}

\item First we restrict our attention to the subset $W'\subset W$ of those $A$ that have rank $3$ as linear maps $E\to S^2F^\vee$ (see section~\ref{app:invariants:invariants}). $W'$ is dense in $W$, because the complement to $W'$ in $W$ can be defined by algebraic equations on matrix elements of $A$: all $3\times3$ minors of the corresponding $3\times10$ matrix should be $0$.

\item Let $D_r\subset\mathbb PS^2F^\vee$ be the subset formed by matrices of rank $\leq r$ (so that $D_4=\mathbb PS^2F^\vee$). $D_r$ are known as \emph{symmetric determinantal varieties}. It is known that $D_r$ is an irreducible algebraic subvariety of $\mathbb PS^2F^\vee$, and the singular locus of $D_r$ is precisely $D_{r-1}$ when $0<r<4$; this is proved in the same way as for general (not necessarily symmetric) matrices in \citep[section II.2]{ACGH1}.  The codimension of $D_r$ is $\binom{5-r}2$ (cf.\ \citep[section 1]{HarTu84}), this can also be proved as in \citep[section II.2]{ACGH1}.

\item $W'/GL(E)$ is the Grassmannian $\operatorname{Gr}(3,S^2F^\vee)=\operatorname{Gr}(2,\mathbb PS^2F^\vee)$ of $2$-planes in $\mathbb PS^2F^\vee$. The equation of $D_3$ is $\det=0$, so the projective quartic curve defined by the equation $\det A=0$ is, by construction, the intersection of the plane $\mathbb P(\operatorname{im}A)\subset \mathbb PS^2F^\vee$ with $D_3$. A transversal intersection of smooth varieties is always smooth, so $\mathrm{discr}(\det(A))$ can only be zero if $\mathbb P(\operatorname{im}A)$ belongs to one of the following two subsets of $\operatorname{Gr}(2,\mathbb PS^2F^\vee)$:

\begin{enumerate}
\item the subset $\mathrm{CH}_0D_2$ consisting of planes whose intersection with $D_3$ contains a singular point of $D_3$ (i.e.\ whose intersection with $D_2$ is non-empty, see above);
\item the subset $\mathrm{CH}_2D_3$, the closure of the subset consisting of planes whose intersection with $D_3$ is non-transversal at some smooth point of $D_3$.
\end{enumerate}

This notation is a particular instance of a more general one: for a variety $X$ embedded into a projective space, $\mathrm{CH}_iX$ denotes what is called the $i$'th \emph{higher associated variety} \citep[section 3.2.E]{GKZ} or the $i$'th \emph{coisotropic variety} \citep[definition 2]{Kohn16} of $X$.

\item If $X$ is irreducible, then $\mathrm{CH}_iX$ is an irreducible subvariety of a Grassmannian, see \citep[proposition 3.2.11]{GKZ} (the proof of irreducibility is similar to that of \citep[proposition 3.2.2]{GKZ} as well).

$\mathrm{CH}_iX$ is a hypersurface if $i\leq\dim X-\operatorname{codim} X^\vee+1$, where $X^\vee$ is the variety dual to $X$ \citep[corollary 6]{Kohn16}. In our case $D_r^\vee\simeq D_{4-r}$ (the proof is analogous to that of  \citep[proposition 1.4.11]{GKZ} for not necessarily symmetric matrices), so it follows from the dimension formulas for $D_r$ given above that $\mathrm{CH}_0D_2$ and $\mathrm{CH}_2D_3$ are irreducible hypersurfaces in $\mathrm{Gr}(2,\mathbb PS^2F^\vee)$.

\item By invariant theory, any irreducible hypersurface $Y$ in the Grassmannian $W'/GL_3$ is the zero set of a homogeneous irreducible $SL(E)$-invariant polynomial $p_Y$ on $W$, and this $p_Y$ is unique up to a constant factor. The degree of $p_Y$ is always divisible by $3$, and the degree of $Y$ can be defined as $\frac13\deg p_Y$ (this is actually the degree of the defining polynomial of $Y$ in Pl\"ucker coordinates, which are cubic in coordinates of $W$). See \citep[propositions 3.1.6 and 3.2.1]{GKZ}.

\item The degrees of $\mathrm{CH}_iD_r$ (when they are hypersurfaces) are known. We indicate where the formulas can be found for the 2 cases we need.
\begin{enumerate}
\item $\deg\mathrm{CH}_0D_2=\deg D_2$ \citep[proposition 3.2.2]{GKZ}, and 
\begin{equation}
\deg D_2=\prod\limits_{\alpha=0}^1\dfrac{\dbinom{4+\alpha}{2-\alpha}}{\dbinom{2\alpha+1}{\alpha}} = \dfrac{\dbinom42}{\dbinom10}\cdot\dfrac{\dbinom51}{\dbinom31}=\frac61\cdot\frac53=10
\end{equation}
by \citep[proposition 12]{HarTu84}.

\item More generally, $\deg\mathrm{CH}_iD_r = \delta\Big(i+\binom{5-r}2,4,r\Big)$, where $\delta$ is the so called ``algebraic degree of semidefinite programming'', see \citep[theorem 2]{JiaStu20}. In particular, $\deg\mathrm{CH}_2D_3=\delta(3,4,3)$. By \citep[theorem 11, point 1]{NRS06} $\delta(3,4,3)=2^2\cdot\binom43=16$.
\end{enumerate}

\item Let $\bar I$ be the $SL(E)$-invariant polynomial on $W$ defining $\mathrm{CH}_0D_2$ and $\bar J$ the one defining $\mathrm{CH}_2D_3$ ($\bar I$ and $\bar J$ are defined up to a constant factor). It follows from the previous points that $\bar I$ and $\bar J$ are irreducible of degrees $\deg\bar I=10\cdot3=30$ and $\deg\bar J=16\cdot3=48$.

Now from Hilbert's Nullstellensatz it follows that the polynomial $\mathrm{discr}\circ\det$ on $W$ is equal to $\bar I^k\bar J^l$ for some $k,l=0,1,2,...$, up to a constant factor. The total degree $\deg(\mathrm{discr}\circ\det)=\deg\mathrm{discr}\cdot\deg\det=27\cdot4=108$, and the equation $30k+48l=108$ has just one solution $k=2,l=1$. The ring of $SL(E)$-invariant polynomials on $W$ is a unique factorization domain \citep[proposition 3.2.1]{GKZ}, so, comparing to Salmon's theorem, we find out that $\bar I=I$ and $\bar J=J$ up to constant factors; in particular, $I$ and $J$ are irreducible.

\end{itemize}

\section{The map from the Siegel upper half-space to the space of nets of quaternary quadrics}\label{app:map}

Here we give an explicit formula for the map $A:H_3\to W$ (from the Siegel upper half-space $H_3$ to the vector space $W$ of symmetric $4\times4$ matrices with $\mathbb C$-linear combinations of three given variables --- say $x_0,x_1,x_2$ --- as entries) such that $A$ is holomorphic and\\
\begin{flalign}\tag{$\ast$}\label{eq:property}
\parbox{0.85\textwidth}{if $\tau\in H_3$ is a period matrix of a non-hyperelliptic Riemann surface, then the Riemann surface $C\subset\mathbb P^2$ defined by the equation $\det A(\tau)=0$ and equipped with the even spin structure induced by this determinantal representation has, for some choice of a symplectic basis of $H_1(C,\mathbb Z)$, period matrix $\tau$ and theta characteristic $\begin{bmatrix}000\\000\end{bmatrix}$.}
\end{flalign}

This $A$ has appeared above in point~\ref{sec:plan:modular-forms} of the plan (section~\ref{sec:plan}). We stress that the explicit form of $A$ is not important for the rest of the paper: we only want $A$ to be holomorphic and to have the property~\eqref{eq:property}. The choice of another $A$ with these 2 properties would only possibly result in a different value of $n$ in section~\ref{sec:derivation:factorization}, all the rest would remain the same.

Our $A$ is a slight modification of the meromorphic map $H_3\to W$ constructed in \citep{DFS17}, see their last corollary 5.3 (6.3 in the preprint). The latter map is denoted $A$ in \citep{DFS17}, but we, on the contrary, use $A$ to denote our modified map and $\tilde A$ to denote the original map of \citep[corollary 5.3]{DFS17}.

Our modification is not strictly necessary from the theoretical point of view, we could have used the original $\tilde A(\tau)$. But the modification makes our formulas, and therefore computer calculations, considerably easier.

In appendix~\ref{app:map:formula} an explicit formula for $A(\tau)$ is written. In appendix~\ref{app:map:difference} we give the original formula of \citep{DFS17}. We also explain informally what modifications we make and what their effect is; the explicit formulas are deferred to appendix~\ref{app:map:transition}.

\subsection{The formula for \texorpdfstring{$A(\tau)$}{\$A(\string\\tau)\$}}\label{app:map:formula}
\begin{equation}\label{eq:map-formula}
A(\tau)=
\begin{pmatrix}
0&*&*&*\\
\theta_{04}\theta_{41}\theta_{50}\theta_{66}\;\beta_{77}\cdot \tilde x & 0   & * & *\\
\theta_{02}\theta_{25}\theta_{34}\theta_{60}\;\beta_{13}\cdot \tilde x & x_2 & 0 & *\\
\theta_{01}\theta_{04}\theta_{10}\theta_{37}\;\beta_{26}\cdot \tilde x & x_1 & x_0 & 0
\end{pmatrix},
\end{equation}
where
\begin{itemize}
\item the elements above the main diagonal are determined by the condition that $A$ be symmetric;
\item $\theta_m=\theta_m(\tau,0)$ is the theta constant with characteristic $m$ (characteristics are encoded by pairs of decimal digits, see appendix~\ref{app:theta});
\item $\theta_{m,i}:=\dfrac\partial{\partial z_i}\Big|_{z=0}\theta_m(\tau,z)$ ($i=0,1,2$);
\item $\beta_m=\begin{pmatrix}\theta_{m,0}&\theta_{m,1}&\theta_{m,2}\end{pmatrix}\begin{pmatrix}
\theta_{35,0}&\theta_{35,1}&\theta_{35,2}\\
\theta_{51,0}&\theta_{51,1}&\theta_{51,2}\\
\theta_{64,0}&\theta_{64,1}&\theta_{64,2}\\
\end{pmatrix}^\vee$, where $\vee$ means the adjoint\footnote{So that $MM^\vee=M^\vee M=(\det M)\mathrm{Id}$ for any square matrix $M$, where $\mathrm{Id}$ means the identity matrix.} matrix;
\item[\vspace{9em}$\bullet$]
$\tilde x
=\begin{pmatrix}
\tilde x_0\\
\tilde x_1\\
\tilde x_2\\
\end{pmatrix}
=\begin{pmatrix}
\theta_{43}\theta_{52}\theta_{75}\theta_{04}\theta_{40}\theta_{67}\theta_{76}x_0\\
\theta_{43}\theta_{52}\theta_{75}\theta_{03}\theta_{12}\theta_{24}\theta_{60}x_1\\
\theta_{04}\theta_{40}\theta_{67}\theta_{76}\theta_{03}\theta_{12}\theta_{24}x_2
\end{pmatrix}.
$
\end{itemize}

\subsection{The difference with the original map \texorpdfstring{$\tilde A(\tau)$}{\$\string\\tilde A(\string\\tau)\$}
}\label{app:map:difference}

Note that the property~\eqref{eq:property} is not affected by the following modifications of $A:H_3\to W$:
\begin{enumerate}
\item Swapping the $i$'th and the $j$'th column followed by swapping of the $i$'th and the $j$'th row. This is equivalent to the conjugation by a certain matrix from $GL_4$.
\item Multiplying $A$ by a meromorphic function $f:H_3\to\mathbb C$ such that on the (open) locus $U_3\subset H_3$ of period matrices of non-hyperelliptic Riemann surfaces $f$ is holomorphic and has no zeros.

For example, we are allowed to multiply by even theta constants: for genus 3, a period matrix $\tau$ is a period matrix of a hyperelliptic Riemann surface if and only if $\theta_m(\tau)=0$ for some even characteristic $m$ \citep[Lemma 11]{Igusa67}.

Even more generally, we may multiply not the whole matrix but just the $i$'th row together with the $i$'th column for some $i$ (this is equivalent to the conjugation by a diagonal $4\times4$ matrix).

\item Linear changes of the independent variables $x_0,x_1,x_2$; the transition matrix may well depend on $\tau$ but should be non-degenerate at any $\tau\in U_3$.
\end{enumerate}

We use this freedom to modify $\tilde A(\tau)$ in the following way:
\begin{enumerate}
\item We apply a linear change of variables to bring our matrix into the form of \citep[example 2.7]{Gizatullin07}.

This allows us to use the formulas for this type of nets only, such formulas are considerably simpler than general ones. In appendix~\ref{app:invariants} we give a self-contained description of all invariants that we use in this paper; if we did not make this modification, then appendix~\ref{app:invariants} would be much longer.

\item $\tilde A$ has a pole at the hyperelliptic locus $H_3\setminus U_3\subset H_3$. We multiply some of the rows and columns by certain holomorphic functions with no zeros on $U_3$ in order to remove this pole.

If we did not make this modification, then we would have to consider separately the numerator and the denominator of some representation of $\tilde A$ as a quotient of 2 holomorphic functions: the left-hand side of~\eqref{eq:factorization} would be undefined if one wrote $\tilde A$ instead of $A$ in~\eqref{eq:factorization}.

\item Another modification does not change the matrix itself, it only changes the formula for it. The original formula for $\tilde A(\tau)$ includes quantities $D(m_1,m_2,m_3)$ called Jacobian Nullwerte; here $m_1,m_2,m_3$ are characteristics. Each Jacobian Nullwert appearing in the original formula for $\tilde A$ is equal, up to a sign, to the product of 5 even theta constants scaled by the factor of $\pi^3$. So we trade Jacobian Nullwerte for theta constants.

This helps us make modification 2 and makes the formula for $A$ simpler.
\end{enumerate}

\subsection{How to get \texorpdfstring{$A(\tau)$}{\$A(\string\\tau)\$} from \texorpdfstring{$\tilde A(\tau)$}{\$\string\\tilde A(\string\\tau)\$}}\label{app:map:transition}

The original formula of \citep[corollary 5.3]{DFS17} is
\begin{equation}\label{eq:map-DFS17}
\tilde A(\tau) = 
\begin{pmatrix}
0 &* &* &* \\
\dfrac{D(31,13,26)}{D(77,31,26)}b_{77}  & 0 & *   & *\\
\dfrac{D(22,13,35)}{D(77,31,26)}b_{64}  & \dfrac{D(22,13,35)}{D(77,46,51)}b_{13} & 0 & *\\
\dfrac{D(77,64,46)}{D(77,31,26)}b_{51}  & \dfrac{D(77,13,31)}{D(77,31,26)}b_{26} & \dfrac{D(64,13,22)}{D(77,31,26)}b_{35} & 0
\end{pmatrix},
\end{equation}
where the notation is the same as in appendix~\ref{app:map:formula} except for the following three points:
\begin{enumerate}
\item The independent variables are $y_0,y_1,y_2$ (and not $x_0,x_1,x_2$).
\item For a characteristic $m$, $b_m = \theta_{m,0}y_0 +\theta_{m,1}y_1 + \theta_{m,2}y_2$.
\item For characteristics $s,t,u$
\begin{equation}
D(s,t,u) := \det\begin{pmatrix}
\theta_{s,0}&\theta_{s,1}&\theta_{s,2}\\
\theta_{t,0}&\theta_{t,1}&\theta_{t,2}\\
\theta_{u,0}&\theta_{u,1}&\theta_{u,2}\\
\end{pmatrix}
\end{equation}
is the Jacobian Nullwert.\footnote{``Nullwert''  (plural ``Nullwerte'') is the German for ``zero value''. Here this refers to the substitution of $z=0$ into derivatives of theta functions $\theta_m(\tau,z)$. Another term for ``theta constant'' $\theta_m(\tau,0)$ is ``Thetanullwert''.}
\end{enumerate}

We modify $\tilde A(\tau)$ via the following steps:

\begin{enumerate}
\item We swap the 1st and the 2nd row, and also the 1st and the 2nd column. Then we multiply the matrix by $D(77,31,26)$. We get
\begin{equation}
\begin{pmatrix}
0 &* &* &* \\
{D(31,13,26)}b_{77}  & 0 & *   & *\\
\dfrac{D(77,31,26)}{D(77,46,51)}D(22,13,35)b_{13}  & {D(22,13,35)}b_{64} & 0 & *\\
{D(77,13,31)}b_{26}  & {D(77,64,46)}b_{51} & {D(64,13,22)}b_{35} & 0
\end{pmatrix}.
\end{equation}

\item We make a linear change of the independent variables: the new independent variables will be
\begin{equation}
\begin{pmatrix}x_0\\x_1\\x_2\end{pmatrix}
=\begin{pmatrix}
D(64,13,22)b_{35}\\
-D(77,64,46)b_{51}\\
-D(22,13,35)b_{64}
\end{pmatrix},\end{equation}
equivalently,
\begin{equation}
\begin{pmatrix}
D(64,13,22)^{-1}x_0\\
-D(77,64,46)^{-1}x_1\\
-D(22,13,35)^{-1}x_2\\
\end{pmatrix}
=J(35,51,64)
\begin{pmatrix}y_0\\y_1\\y_2 \end{pmatrix}
\end{equation}
with
\begin{equation}
J(35,51,64):=\begin{pmatrix}
\theta_{35,0}&\theta_{35,1}&\theta_{35,2}\\
\theta_{51,0}&\theta_{51,1}&\theta_{51,2}\\
\theta_{64,0}&\theta_{64,1}&\theta_{64,2}\\
\end{pmatrix},
\end{equation}
equivalently,
\begin{equation}
\begin{pmatrix}y_0\\y_1\\y_2 \end{pmatrix}
=D(35,51,64)^{-1}
J(35,51,64)^\vee
\begin{pmatrix}
D(64,13,22)^{-1}x_0\\
-D(77,64,46)^{-1}x_1\\
-D(22,13,35)^{-1}x_2
\end{pmatrix}.
\end{equation}

We also multiply the 1st row and the 1st column by \begin{equation}
D(35,51,64)D(64,13,22)D(77,64,46)D(22,13,35)D(77,46,51).
\end{equation}
 Thus we bring our matrix to the form

\begin{equation}\label{eq:map-formula-non-reduced}
\begin{pmatrix}
0&*&*&*\\
D(31,13,26)D(77,46,51)\tilde\beta_{77} & 0 & * & *\\
D(77,31,26)D(22,13,35)\tilde\beta_{13} & -x_2 & 0 & *\\
D(77,13,31)D(77,46,51)\tilde\beta_{26} & -x_1 & x_0 & 0
\end{pmatrix},
\end{equation}
where
\begin{equation}
\tilde\beta_m
= \begin{pmatrix}\theta_{m,0}&\theta_{m,1}&\theta_{m,2}\end{pmatrix}
J(35,51,64)^\vee
\begin{pmatrix}
D(77,64,46)D(22,13,35)x_0\\
-D(64,13,22)D(22,13,35)x_1\\
-D(64,13,22)D(77,64,46)x_2
\end{pmatrix}
\end{equation}

\item If $\{m_1,m_2,m_3\}$ is an azygetic set of 3 pairwise different odd characteristics, then there is a unique set of 5 pairwise different even characteristics $\{m_4,m_5,m_6,m_7,m_8\}$ such that the set $\{m_1,m_2,...,m_8\}$ is azygetic. Moreover, $D(m_1,m_2,m_3)=\pm\pi^3\prod\limits_{i=4}^8\theta_{m_i}$. See \citep{Igusa80}.

Explicitly, for the Jacobian Nullwerte appearing in~\eqref{eq:map-DFS17} the formulas are given in table~\ref{tab} (which was filled in with the help of a computer).
\begin{table}
\[
\begin{array}{|c|c|c|}
\hline
m_1,m_2,m_3 & m_4,m_5,m_6,m_7,m_8 & \text{The sign in }D(m_1,m_2,m_3)=\pm\pi^3\prod\limits_{i=4}^8\theta_{m_i}\\
\hline
22,13,35    & 00,43,52,60,75      & -\\
31,13,26    & 00,41,50,66,73      & +\\
64,13,22    & 00,03,12,24,60      & +\\
77,13,31    & 00,01,10,37,73      & -\\
77,31,26    & 00,02,25,34,73      & +\\
77,46,51    & 00,04,43,52,75      & +\\
77,64,46    & 00,04,40,67,76      & -\\
\hline
\end{array}
\]
\caption{\label{tab}Jacobian Nullwerte as products of even theta constants.}
\end{table}

Substituting this into~\eqref{eq:map-formula-non-reduced} and dividing the 1st row and the 1st column by their common factor $\theta_{00}^4\theta_{43}\theta_{52}\theta_{60}\theta_{73}\theta_{75}$, we get

\begin{equation}
\begin{pmatrix}
0&*&*&*\\
\theta_{04}\theta_{41}\theta_{50}\theta_{66}\;\beta_{77}\cdot \tilde x & 0   & -x_2 & -x_1\\
-\theta_{02}\theta_{25}\theta_{34}\theta_{60}\;\beta_{13}\cdot \tilde x & -x_2 & 0 & x_0\\
-\theta_{01}\theta_{04}\theta_{10}\theta_{37}\;\beta_{26}\cdot \tilde x & -x_1 & x_0 & 0
\end{pmatrix}.
\end{equation}

Now we multiply the 1st row and the 1st column by $-1$, then do the same with the 2nd row and the 2nd column and get~\eqref{eq:map-formula}.

\end{enumerate}

\section{The explicit formula for the invariant \texorpdfstring{$Q'$}{Q'}}\label{app:Qprime}
As a polynomial in $a,b,c,e,f,g,p,q,r$, the invariant $Q'$ (see appendix~\ref{app:invariants:Qprime}) has the following explicit form (this formula was obtained with the help of a computer):\\[1em]
$
Q'=\input{Qprime}
.
$

{
\acknowledgments


We are grateful to M.~Gizatullin, A.~Kuznetsov, A.~Litvinov, A.~Losev, A.~Marshakov, A.~Morozov, A.~Rosly and G.~Shabat for useful discussions, and to the anonymous referee for comments and suggestions.


I.F.\ and P.D.-B.\ gratefully acknowledge support from the Basic Research Program of the National Research University Higher School of Economics. A.S.\ gratefully acknowledges support from the Ministry of Science and Higher Education of the Russian Federation (agreement no.\ 075-03-2025-662).

Computer experiments were conducted with the use of the system \texttt{SageMath} \citep{sagemath}; we used the package \texttt{RiemannTheta} \citep{riemanntheta} to compute theta functions.

}

\bibliographystyle{JHEP}
\bibliography{superstring-measure-genus-3}

\end{document}

%% file: Qprime.tex
-3 c^6 f^3 + 9 b c^5 f^2 g - 9 c^5 e f^2 g - 9 b^2 c^4 f g^2 + 18 b c^4 e f g^2 - 9 c^4 e^2 f g^2 + 9 a c^4 f^2 g^2 + 3 b^3 c^3 g^3 - 9 b^2 c^3 e g^3 + 9 b c^3 e^2 g^3 - 3 c^3 e^3 g^3 - 18 a b c^3 f g^3 + 18 a c^3 e f g^3 + 9 a b^2 c^2 g^4 - 18 a b c^2 e g^4 + 9 a c^2 e^2 g^4 - 9 a^2 c^2 f g^4 + 9 a^2 b c g^5 - 9 a^2 c e g^5 + 3 a^3 g^6 + 9 b c^4 f^2 g p + 9 c^4 e f^2 g p - 18 b^2 c^3 f g^2 p + 18 c^3 e^2 f g^2 p + 9 b^3 c^2 g^3 p - 9 b^2 c^2 e g^3 p - 9 b c^2 e^2 g^3 p + 9 c^2 e^3 g^3 p - 18 a b c^2 f g^3 p - 18 a c^2 e f g^3 p + 18 a b^2 c g^4 p - 18 a c e^2 g^4 p + 9 a^2 b g^5 p + 9 a^2 e g^5 p + 9 c^4 f^3 p^2 - 22 b c^3 f^2 g p^2 + 10 c^3 e f^2 g p^2 + 6 b^2 c^2 f g^2 p^2 - 24 b c^2 e f g^2 p^2 - 14 a c^2 f^2 g^2 p^2 + 7 b^3 c g^3 p^2 + 7 b^2 c e g^3 p^2 - 5 b c e^2 g^3 p^2 - 9 c e^3 g^3 p^2 + 10 a b c f g^3 p^2 - 22 a c e f g^3 p^2 + 11 a b^2 g^4 p^2 + 16 a b e g^4 p^2 + 9 a e^2 g^4 p^2 + 11 a^2 f g^4 p^2 - 10 b c^2 f^2 g p^3 - 10 c^2 e f^2 g p^3 + 10 b^2 c f g^2 p^3 - 4 b c e f g^2 p^3 - 18 c e^2 f g^2 p^3 - 4 a c f^2 g^2 p^3 + 5 b^3 g^3 p^3 + 7 b^2 e g^3 p^3 + 9 b e^2 g^3 p^3 + 3 e^3 g^3 p^3 + 22 a b f g^3 p^3 + 18 a e f g^3 p^3 - 9 c^2 f^3 p^4 + 5 b c f^2 g p^4 - 9 c e f^2 g p^4 + 11 b^2 f g^2 p^4 + 18 b e f g^2 p^4 + 9 e^2 f g^2 p^4 + 9 a f^2 g^2 p^4 + 9 b f^2 g p^5 + 9 e f^2 g p^5 + 3 f^3 p^6 - 9 b c^5 f^2 q - 9 c^5 e f^2 q + 18 b^2 c^4 f g q - 18 c^4 e^2 f g q - 9 b^3 c^3 g^2 q + 9 b^2 c^3 e g^2 q + 9 b c^3 e^2 g^2 q - 9 c^3 e^3 g^2 q + 18 a b c^3 f g^2 q + 18 a c^3 e f g^2 q - 18 a b^2 c^2 g^3 q + 18 a c^2 e^2 g^3 q - 9 a^2 b c g^4 q - 9 a^2 c e g^4 q + 11 b c^4 f^2 p q - 5 c^4 e f^2 p q - 6 b^2 c^3 f g p q + 56 b c^3 e f g p q + 4 c^3 e^2 f g p q + 2 a c^3 f^2 g p q - 5 b^3 c^2 g^2 p q - 33 b^2 c^2 e g^2 p q + 33 b c^2 e^2 g^2 p q + 5 c^2 e^3 g^2 p q - 16 a b c^2 f g^2 p q + 16 a c^2 e f g^2 p q - 4 a b^2 c g^3 p q - 56 a b c e g^3 p q + 6 a c e^2 g^3 p q - 2 a^2 c f g^3 p q + 5 a^2 b g^4 p q - 11 a^2 e g^4 p q + 12 b c^3 f^2 p^2 q + 10 c^3 e f^2 p^2 q - 26 b^2 c^2 f g p^2 q + 24 c^2 e^2 f g p^2 q + 7 b^3 c g^2 p^2 q - 43 b^2 c e g^2 p^2 q - 33 b c e^2 g^2 p^2 q + 9 c e^3 g^2 p^2 q - 2 a b c f g^2 p^2 q - 12 a c e f g^2 p^2 q + 10 a b^2 g^3 p^2 q - 2 a b e g^3 p^2 q - 18 a e^2 g^3 p^2 q - 4 a^2 f g^3 p^2 q - 12 b c^2 f^2 p^3 q + 22 c^2 e f^2 p^3 q - 2 b^2 c f g p^3 q - 56 b c e f g p^3 q - 2 a c f^2 g p^3 q + 7 b^3 g^2 p^3 q + 7 b^2 e g^2 p^3 q - 9 b e^2 g^2 p^3 q - 9 e^3 g^2 p^3 q + 12 a b f g^2 p^3 q - 18 a e f g^2 p^3 q - 11 b c f^2 p^4 q - 9 c e f^2 p^4 q + 16 b^2 f g p^4 q - 18 e^2 f g p^4 q + 9 b f^2 p^5 q - 9 e f^2 p^5 q - 9 b^2 c^4 f q^2 - 16 b c^4 e f q^2 - 11 c^4 e^2 f q^2 - 11 a c^4 f^2 q^2 + 9 b^3 c^3 g q^2 + 5 b^2 c^3 e g q^2 - 7 b c^3 e^2 g q^2 - 7 c^3 e^3 g q^2 + 22 a b c^3 f g q^2 - 10 a c^3 e f g q^2 + 24 a b c^2 e g^2 q^2 - 6 a c^2 e^2 g^2 q^2 + 14 a^2 c^2 f g^2 q^2 - 10 a^2 b c g^3 q^2 + 22 a^2 c e g^3 q^2 - 9 a^3 g^4 q^2 + 18 b^2 c^3 f p q^2 + 2 b c^3 e f p q^2 - 10 c^3 e^2 f p q^2 + 4 a c^3 f^2 p q^2 - 9 b^3 c^2 g p q^2 + 33 b^2 c^2 e g p q^2 + 43 b c^2 e^2 g p q^2 - 7 c^2 e^3 g p q^2 + 12 a b c^2 f g p q^2 + 2 a c^2 e f g p q^2 - 24 a b^2 c g^2 p q^2 + 26 a c e^2 g^2 p q^2 - 10 a^2 b g^3 p q^2 - 12 a^2 e g^3 p q^2 + 26 b c^2 e f p^2 q^2 - 6 c^2 e^2 f p^2 q^2 + 16 a c^2 f^2 p^2 q^2 - 9 b^3 c g p^2 q^2 - 33 b^2 c e g p^2 q^2 + 33 b c e^2 g p^2 q^2 + 9 c e^3 g p^2 q^2 - 18 a b c f g p^2 q^2 + 18 a c e f g p^2 q^2 + 6 a b^2 g^2 p^2 q^2 - 26 a b e g^2 p^2 q^2 - 16 a^2 f g^2 p^2 q^2 - 18 b^2 c f p^3 q^2 + 6 b c e f p^3 q^2 + 18 c e^2 f p^3 q^2 + 9 b^3 g p^3 q^2 - 5 b^2 e g p^3 q^2 - 9 b e^2 g p^3 q^2 + 9 e^3 g p^3 q^2 - 16 a b f g p^3 q^2 - 18 a e f g p^3 q^2 + 9 b^2 f p^4 q^2 - 18 b e f p^4 q^2 + 9 e^2 f p^4 q^2 - 9 a f^2 p^4 q^2 - 3 b^3 c^3 q^3 - 9 b^2 c^3 e q^3 - 7 b c^3 e^2 q^3 - 5 c^3 e^3 q^3 - 18 a b c^3 f q^3 - 22 a c^3 e f q^3 + 18 a b^2 c^2 g q^3 + 4 a b c^2 e g q^3 - 10 a c^2 e^2 g q^3 + 4 a^2 c^2 f g q^3 + 10 a^2 b c g^2 q^3 + 10 a^2 c e g^2 q^3 + 9 b^3 c^2 p q^3 + 9 b^2 c^2 e p q^3 - 7 b c^2 e^2 p q^3 - 7 c^2 e^3 p q^3 + 18 a b c^2 f p q^3 - 12 a c^2 e f p q^3 + 56 a b c e g p q^3 + 2 a c e^2 g p q^3 + 2 a^2 c f g p q^3 - 22 a^2 b g^2 p q^3 + 12 a^2 e g^2 p q^3 - 9 b^3 c p^2 q^3 + 9 b^2 c e p^2 q^3 + 5 b c e^2 p^2 q^3 - 9 c e^3 p^2 q^3 + 18 a b c f p^2 q^3 + 16 a c e f p^2 q^3 - 18 a b^2 g p^2 q^3 - 6 a b e g p^2 q^3 + 18 a e^2 g p^2 q^3 + 3 b^3 p^3 q^3 - 9 b^2 e p^3 q^3 + 9 b e^2 p^3 q^3 - 3 e^3 p^3 q^3 - 18 a b f p^3 q^3 + 18 a e f p^3 q^3 - 9 a b^2 c^2 q^4 - 18 a b c^2 e q^4 - 11 a c^2 e^2 q^4 - 9 a^2 c^2 f q^4 + 9 a^2 b c g q^4 - 5 a^2 c e g q^4 + 9 a^3 g^2 q^4 + 18 a b^2 c p q^4 - 16 a c e^2 p q^4 + 9 a^2 b g p q^4 + 11 a^2 e g p q^4 - 9 a b^2 p^2 q^4 + 18 a b e p^2 q^4 - 9 a e^2 p^2 q^4 + 9 a^2 f p^2 q^4 - 9 a^2 b c q^5 - 9 a^2 c e q^5 + 9 a^2 b p q^5 - 9 a^2 e p q^5 - 3 a^3 q^6 + 9 b^2 c^4 f^2 r - 9 c^4 e^2 f^2 r - 18 b^3 c^3 f g r + 18 b^2 c^3 e f g r + 18 b c^3 e^2 f g r - 18 c^3 e^3 f g r + 9 b^4 c^2 g^2 r - 18 b^3 c^2 e g^2 r + 18 b c^2 e^3 g^2 r - 9 c^2 e^4 g^2 r - 18 a b^2 c^2 f g^2 r + 18 a c^2 e^2 f g^2 r + 18 a b^3 c g^3 r - 18 a b^2 c e g^3 r - 18 a b c e^2 g^3 r + 18 a c e^3 g^3 r + 9 a^2 b^2 g^4 r - 9 a^2 e^2 g^4 r + 2 b c^3 e f^2 p r + 4 c^3 e^2 f^2 p r - 16 b^3 c^2 f g p r - 18 b^2 c^2 e f g p r + 12 b c^2 e^2 f g p r + 22 c^2 e^3 f g p r + 6 a b c^2 f^2 g p r + 8 a c^2 e f^2 g p r + 16 b^4 c g^2 p r - 2 b^3 c e g^2 p r - 26 b^2 c e^2 g^2 p r - 6 b c e^3 g^2 p r + 18 c e^4 g^2 p r - 10 a b^2 c f g^2 p r - 20 a b c e f g^2 p r + 22 a b^3 g^3 p r + 12 a b^2 e g^3 p r - 16 a b e^2 g^3 p r - 18 a e^3 g^3 p r + 10 a^2 b f g^3 p r - 16 b^2 c^2 f^2 p^2 r + 14 c^2 e^2 f^2 p^2 r + 12 b^3 c f g p^2 r - 2 b^2 c e f g p^2 r - 16 b c e^2 f g p^2 r + 18 c e^3 f g p^2 r - 32 a b c f^2 g p^2 r - 8 a c e f^2 g p^2 r + 11 b^4 g^2 p^2 r + 10 b^3 e g^2 p^2 r + 6 b^2 e^2 g^2 p^2 r - 18 b e^3 g^2 p^2 r - 9 e^4 g^2 p^2 r + 34 a b^2 f g^2 p^2 r - 10 a b e f g^2 p^2 r - 18 a e^2 f g^2 p^2 r + 8 a^2 f^2 g^2 p^2 r - 4 b^2 c f^2 p^3 r - 2 b c e f^2 p^3 r + 22 b^3 f g p^3 r + 10 b^2 e f g p^3 r - 18 b e^2 f g p^3 r - 18 e^3 f g p^3 r + 10 a b f^2 g p^3 r + 11 b^2 f^2 p^4 r - 9 e^2 f^2 p^4 r + 18 b^3 c^3 f q r + 16 b^2 c^3 e f q r - 12 b c^3 e^2 f q r - 22 c^3 e^3 f q r - 10 a c^3 e f^2 q r - 18 b^4 c^2 g q r + 6 b^3 c^2 e g q r + 26 b^2 c^2 e^2 g q r + 2 b c^2 e^3 g q r - 16 c^2 e^4 g q r + 20 a b c^2 e f g q r + 10 a c^2 e^2 f g q r - 22 a b^3 c g^2 q r - 12 a b^2 c e g^2 q r + 18 a b c e^2 g^2 q r + 16 a c e^3 g^2 q r - 8 a^2 b c f g^2 q r - 6 a^2 c e f g^2 q r - 4 a^2 b^2 g^3 q r - 2 a^2 b e g^3 q r - 18 b^3 c^2 f p q r + 18 b^2 c^2 e f p q r + 2 b c^2 e^2 f p q r - 10 c^2 e^3 f p q r + 8 a b c^2 f^2 p q r + 32 a c^2 e f^2 p q r - 56 b^3 c e g p q r + 56 b c e^3 g p q r - 20 a b^2 c f g p q r + 20 a c e^2 f g p q r + 10 a b^3 g^2 p q r - 2 a b^2 e g^2 p q r - 18 a b e^2 g^2 p q r + 18 a e^3 g^2 p q r - 32 a^2 b f g^2 p q r - 8 a^2 e f g^2 p q r - 18 b^3 c f p^2 q r - 12 b^2 c e f p^2 q r + 16 b c e^2 f p^2 q r + 18 c e^3 f p^2 q r - 8 a b c f^2 p^2 q r - 6 a c e f^2 p^2 q r + 18 b^4 g p^2 q r - 4 b^3 e g p^2 q r - 24 b^2 e^2 g p^2 q r + 18 e^4 g p^2 q r - 10 a b^2 f g p^2 q r - 20 a b e f g p^2 q r + 18 b^3 f p^3 q r - 22 b^2 e f p^3 q r - 18 b e^2 f p^3 q r + 18 e^3 f p^3 q r + 9 b^4 c^2 q^2 r + 18 b^3 c^2 e q^2 r - 6 b^2 c^2 e^2 q^2 r - 10 b c^2 e^3 q^2 r - 11 c^2 e^4 q^2 r + 18 a b^2 c^2 f q^2 r + 10 a b c^2 e f q^2 r - 34 a c^2 e^2 f q^2 r - 8 a^2 c^2 f^2 q^2 r - 18 a b^3 c g q^2 r + 16 a b^2 c e g q^2 r + 2 a b c e^2 g q^2 r - 12 a c e^3 g q^2 r + 8 a^2 b c f g q^2 r + 32 a^2 c e f g q^2 r - 14 a^2 b^2 g^2 q^2 r + 16 a^2 e^2 g^2 q^2 r - 18 b^4 c p q^2 r + 24 b^2 c e^2 p q^2 r + 4 b c e^3 p q^2 r - 18 c e^4 p q^2 r + 20 a b c e f p q^2 r + 10 a c e^2 f p q^2 r - 18 a b^3 g p q^2 r - 16 a b^2 e g p q^2 r + 12 a b e^2 g p q^2 r + 18 a e^3 g p q^2 r + 6 a^2 b f g p q^2 r + 8 a^2 e f g p q^2 r + 9 b^4 p^2 q^2 r - 18 b^3 e p^2 q^2 r + 18 b e^3 p^2 q^2 r - 9 e^4 p^2 q^2 r - 18 a b^2 f p^2 q^2 r + 18 a e^2 f p^2 q^2 r + 18 a b^3 c q^3 r + 18 a b^2 c e q^3 r - 10 a b c e^2 q^3 r - 22 a c e^3 q^3 r - 10 a^2 c e f q^3 r + 2 a^2 b e g q^3 r + 4 a^2 e^2 g q^3 r - 18 a b^3 p q^3 r + 18 a b^2 e p q^3 r + 22 a b e^2 p q^3 r - 18 a e^3 p q^3 r + 9 a^2 b^2 q^4 r - 11 a^2 e^2 q^4 r - 9 b^4 c^2 f r^2 + 16 b^2 c^2 e^2 f r^2 + 4 b c^2 e^3 f r^2 - 11 c^2 e^4 f r^2 - 8 a c^2 e^2 f^2 r^2 + 9 b^5 c g r^2 - 11 b^4 c e g r^2 - 12 b^3 c e^2 g r^2 + 12 b^2 c e^3 g r^2 + 11 b c e^4 g r^2 - 9 c e^5 g r^2 - 8 a b^2 c e f g r^2 + 8 a b c e^2 f g r^2 + 11 a b^4 g^2 r^2 - 4 a b^3 e g^2 r^2 - 16 a b^2 e^2 g^2 r^2 + 9 a e^4 g^2 r^2 + 8 a^2 b^2 f g^2 r^2 - 2 b^3 c e f p r^2 + 2 b c e^3 f p r^2 + 9 b^5 g p r^2 + 5 b^4 e g p r^2 - 10 b^3 e^2 g p r^2 - 22 b^2 e^3 g p r^2 + 9 b e^4 g p r^2 + 9 e^5 g p r^2 + 10 a b^3 f g p r^2 - 32 a b^2 e f g p r^2 + 6 a b e^2 f g p r^2 + 32 a^2 b f^2 g p r^2 + 9 b^4 f p^2 r^2 - 4 b^3 e f p^2 r^2 - 14 b^2 e^2 f p^2 r^2 + 9 e^4 f p^2 r^2 + 8 a b^2 f^2 p^2 r^2 - 9 b^5 c q r^2 - 9 b^4 c e q r^2 + 22 b^3 c e^2 q r^2 + 10 b^2 c e^3 q r^2 - 5 b c e^4 q r^2 - 9 c e^5 q r^2 - 6 a b^2 c e f q r^2 + 32 a b c e^2 f q r^2 - 10 a c e^3 f q r^2 - 32 a^2 c e f^2 q r^2 - 2 a b^3 e g q r^2 + 2 a b e^3 g q r^2 + 9 b^5 p q r^2 - 9 b^4 e p q r^2 - 10 b^3 e^2 p q r^2 + 10 b^2 e^3 p q r^2 + 9 b e^4 p q r^2 - 9 e^5 p q r^2 - 8 a b^2 e f p q r^2 + 8 a b e^2 f p q r^2 - 9 a b^4 q^2 r^2 + 14 a b^2 e^2 q^2 r^2 + 4 a b e^3 q^2 r^2 - 9 a e^4 q^2 r^2 - 8 a^2 e^2 f q^2 r^2 + 3 b^6 r^3 - 9 b^4 e^2 r^3 + 9 b^2 e^4 r^3 - 3 e^6 r^3